\documentclass[dvipsnames]{aa}  
\usepackage{graphicx}
\graphicspath{{figures/}} 
\DeclareGraphicsExtensions{.pdf,.png,.jpg}
\usepackage{placeins}
\usepackage{float}
\usepackage{xcolor}
\usepackage{array}

\usepackage{lmodern}
\usepackage{amsmath, amssymb}
\usepackage{pgfplots}
\pgfplotsset{compat=1.18}

\usepackage{import}

\usepackage{aas_macros}
\usepackage{natbib}

\usepackage{url}
\usepackage[colorlinks=True,
            linkcolor=gray,         
            citecolor=RoyalBlue,    
            filecolor=gray,         
            urlcolor=gray           
           ]{hyperref}

\usepackage[nameinlink,capitalize]{cleveref}
\crefformat{equation}{Eq.~#2#1#3}
\crefname{section}{Sect.}{Sects.} 
\usepackage[switch]{lineno} 
\usepackage[nolist,nohyperlinks]{acronym}

\usepackage{diagbox}
\usepackage{multirow}
\usepackage{booktabs}

\usepackage{xspace}
\usepackage[normalem]{ulem}

\newcommand{\answer}[1]{{#1}}

\newcommand{\numb}[1]{\textcolor{black}{#1}}

\newcommand{\ssodnet}{\texttt{SsODNet}\xspace}
\newcommand{\rocks}{\texttt{rocks}\xspace}

\newcommand{\nGaiaTot}{60,518} 
\newcommand{\nGaiaRawAllBands}{34,577} 
\newcommand{\nGaiaBestQ}{9,926} 
\newcommand{\nGaiaPv}{41,430} 
\newcommand{\nClassif}{14,042} 
\newcommand{\ratioclassif}{three} 

\begin{document} 

\let\textbf\relax

   \title{Taxonomy of  14042 asteroids from Gaia DR3 reflectance spectra
   PREPRINT}

   \author{F. Tinaut-Ruano
          \inst{1}
          \and
          B. Carry\inst{1}
          \and
          M. Galinier\inst{2}
          \and
          M. Mahlke\inst{3}
          \and
          A. Sergeyev\inst{1}
          }

   \institute{Universit\'e C\^ote d'Azur, Observatoire de la C\^ote d'Azur, CNRS, Laboratoire Lagrange, France\\
     \email{ftinautruano@gmail.com}
      \and
      INAF-IAPS, Institute for Space Astrophysics and Planetology, via del Fosso del Cavaliere, 100, 00133, Roma, Italy
      \and
      Université Marie et Louis Pasteur, CNRS, Institut UTINAM (UMR 6213), équipe Astro, F-25000 Besançon, France
     }

   \date{Received 29/01/2026; accepted 27/04/2026}

 
  \abstract
   { Asteroid reflectance spectra provide key constraints on surface composition. Gaia Data Release 3 (DR3) enables the study of \numb{\nGaiaTot} asteroids through near-ultraviolet (NUV) to visible reflectance spectra.}
   {We aim to classify asteroids using Gaia DR3 spectra and provide a homogeneous taxonomical framework. Owing to systematics affecting Gaia DR3 data, direct comparison with previous taxonomies has to be taken with caution;
   therefore, we developed a classification scheme tailored to Gaia and linked the resulting taxa to established classes.}
   { We selected the highest-quality spectra using Gaia DR3 quality flags and apply uncertainty thresholds to mitigate spectral artifacts, retaining over one third of the original sample at the least noisy wavelengths. To improve compositional discrimination, we included albedo information, reducing the final sample to about
   one fourth of its initial size. We then iteratively applied dimensionality reduction and clustering techniques to identify the spectral taxa.
   }
   {
   We classified \numb{\nClassif} asteroids into 13 taxonomic classes: A, B, C, D, E, F, G, K, L, M, P, S, and V, representing an increase of \numb{\ratioclassif} compared to the number of objects classified in previous spectral classifications. The largest relative increase is found for the K class. The inclusion of NUV wavelengths allows the separation of B and F types within the C-complex and facilitates the identification of G types. The dynamical distribution follows expected trends, with S types dominating the inner and middle Main Belt, C-complex asteroids prevalent in the outer Main Belt, and D types beyond.
   }
   {We present a taxonomical classification of \numb{\nClassif} asteroids based on Gaia DR3 reflectance spectra. NUV coverage is critical for disentangling primitive classes within the C-complex. Although artifacts in Gaia DR3 require caution when comparing median spectra with other datasets, this classification provides a robust reference for future Gaia releases, with larger observed samples.
   } 

   \maketitle

\section{Introduction}\label{sec:int}

Asteroid reflectance spectra and spectrophotometry provide information on the composition of asteroid surfaces, and on the processes that modify their properties, such as space weathering \citep{2015aste.book...43R}. Historically, the use of photoelectric detectors (or photometers), more sensitive at blue wavelengths (i.e., < 0.5 $\mu$m), and the development of the standard $UBV$ photometric system \citep{1951ApJ...114..522J}, led to the appearance of the first asteroid taxonomies in the 1970s \citep{1973BAAS....5..388Z,1975Icar...25..104C} which contained information at blue-visible wavelengths (near-UV or NUV hereafter, from 0.3 to 0.5\,$\mu$m). The introduction of charge-coupled devices (CCDs) in astronomy in the 1990s and later on, contributed to the ``loss'' of NUV information, as CCDs are less sensitive at those wavelengths. Therefore, the large majority of the modern spectroscopic and spectrophotometric surveys cover the wavelength range from $\sim$0.5 $\mu$m up to 2.5 $\mu$m.
Nevertheless, there are some exceptions. One of the first large surveys with information in the NUV is the Eight Color Asteroid Survey (ECAS, \citealt{1985Icar...61..355Z}). They provided photometry in eight broad band filters between 0.34 to 1.04 $\mu$m for 589 minor planets, including two filters below 0.45 $\mu$m. These observations were used to develop a new taxonomy
\citep[see][]{1984PhDT.........3T}.
More recent catalogues, 
like the Sloan Digital Sky Survey (SDSS) Solar System object catalogue \citep{2021A&A...652A..59S},
or the Solar System Objects observations from the SkyMapper Southern Survey \citep{2022A&A...658A.109S}
include NUV photometry for \numb{~800,000} and \numb{205,515} objects, respectively. However the spectral resolution in this surveys is limited having \numb{five} and \numb{six} filters between 0.3 and 1.1 $\mu$m.
On the other hand, the Moving Objects Observed from Javalambre (MOOJa) catalogue from the J-PLUS survey \citep{2021A&A...655A..47M}, has a larger spectral resolution of \numb{12} filters but a limited number of asteroids observed: \numb{3122}.
The Gaia space mission \citep{prusti2016}
offers in its Data Release 3 (DR3) a Solar system survey with different data products \citep{tanga_dr3},
including the reflectance spectra of \numb{\nGaiaTot} objects binned in 16 wavelengths bands between 0.352 and 1.056\,$\mu$m \citep{2023A&A...674A..35G}. It offers a unique opportunity to explore a large amount of asteroids with a moderate spectral resolution and avoiding the atmospheric absorption in the UV.\\

Some systematic effects are, however, present in the Gaia DR3 spectra. These systematics are:
(i) The selection of solar analogs by the Gaia team included a large number of the most commonly used stars by the planetary science community \citep{2023A&A...674A..35G}. Such stars can be used to get asteroid reflectance spectra beyond 0.5 $\mu$m without any problems. However, as it has been shown in \cite{2022A&A...664A.107T} and \cite{2023A&A...669L..14T}, they do not have a Sun-like spectral behavior at wavelengths below 0.5 $\mu$m, introducing an artificial reddening in that region.
(ii) Gaia reflectance spectra show a reddening between 0.7 and 1 $\mu$m as different authors found when comparing Gaia DR3 with other ground-based data sources \citep{2023MT........B, 2023A&A...671A..40G, 2026MNRAS.545f2052M}\answer{, in the first work, Balossi found that this effect is taxonomy dependent}.
(iii) The three bluest and reddest reflectance points in Gaia DR3 spectra are often overestimated as \cite{2023A&A...674A..35G} described.
(iv) An artificial band can be found around 0.65 $\mu$m, due to the merging of the blue and red parts of the spectra in the production of the mean reflectance spectra \citep{2023A&A...674A..35G, 2023A&A...671A..40G, 2024A&A...686A..76T}.
(v) For the V-types, a shift in band center and band depth is detected by \cite{2023Oszkiewicz_basaltic_Gaia}. Comparisons to other spectral databases has to be done with caution due to those artifacts, making the classification of their spectra more difficult. Thus, we decided to classify the reflectance spectra of asteroids in Gaia DR3
by following the steps of building a new taxonomy based on Gaia spectra and make it compatible with previous taxonomies.\\

In this article, we aim to classify a large fraction of the Gaia DR3 asteroids' reflectance spectra catalog through variance driven dimensional reduction and clustering methods. \answer{The base dataset, the cleaning and other methods are presented in \Cref{sec:met}. The iterative process and the decision tree are extensively described in section \Cref{sec:res}. We show the results in \Cref{sec:res2}. Finally, the conclusions are presented in \Cref{sec:con}}.\\

\section{Methodology}\label{sec:met} 

We describe here the selection of the data set, the tools we used, and the iterative clustering algorithm.

\subsection{Selection of the data set}\label{ssec:sam}
To build a taxonomy through variance driven dimensional reduction and clustering methods, we needed to use a sample whose variance is dominated by the mineralogical properties of the asteroids, and not by the noise or systematics present in the data. In this section we show how we selected the best quality data from the Gaia DR3 reflectance spectra catalog.\\

\subsubsection{Gaia reflectance spectra}
Among the Gaia DR3 products for Solar System Objects (SSOs), 
the mean reflectance spectra of Solar system small objects provided in the Gaia DR3 span the wavelength range from 374 to 1034 nm in 16 bands, and are normalised at 550 nm. To each band was assigned a quality flag, indicative of the estimated quality of the band:  0 for good quality, 1 for lower quality and 2 for bad quality.
\citep[see][for more information]{2023A&A...674A..35G}. From the original number of objects provided by Gaia DR3 (\nGaiaTot), more than half (\nGaiaRawAllBands) have data published for all 16 bands in the original dataset. However, less than a sixth (\nGaiaBestQ) have all the bands observed with the best quality flag. In this work, we used only those bands with the best quality flag.\\

Regardless of the quality flag, some bands have some systematic issues. Among those systematics, the reddening due to the stellar analog selection has a known correction \citep{2023A&A...669L..14T} that we applied to the whole dataset. The increase in slope between 0.7 and 1 micron, which renders the comparison of Gaia DR3 with literature spectra complex, seems to be \answer{affecting different taxons in a different way, but consistently inside the taxon.} \citep{2023MT........B, 2023Oszkiewicz_basaltic_Gaia, 2023A&A...671A..40G,2026MNRAS.545f2052M}. Thus we do not have to correct or mask it as it is consistent with the mineralogy, not affecting the clustering methods. However, the overestimation of the reflectance through the first and last three bands is dependent on the estimated error of each band, thus, a low signal asteroid will have a larger change in slope than a bright asteroid, without implications on the mineralogical properties. This is affecting dimensionality reduction algorithms, and spreading the clusters in the feature space. We show (Fig. \ref{fig:dat:SNR_cut}) an example of three density distributions of the spectral slope between the two first, two central and two last bands, versus the uncertainty of the shortest wavelength band. \answer{We selected the threshold based on the error ($T_\epsilon$) as the value of $\epsilon^{-1}$ (being $\epsilon$ the error of the reflectance per wavelength computed by Gaia DPAC)  where the slope trend begins to change its quasi-constant regime for good signal asteroids. In Fig. \ref{fig:dat:SNR_cut} we plotted the selected $T_\epsilon$ for 3 different wavelengths with a pointed line. We tried to find a compromise between the quality of the data, that will lead to a more accurate cluster selection, and the size of the dataset, since a large dataset is needed to sample the less numerous taxa.  We kept in the samples the objects at every wavelength which $\epsilon^{-1}$ > $T_\epsilon$, we call this number $N_\epsilon$. This number and the used values for the threshold per wavelength are in Table \ref{tab:data:number}}.\\

\begin{table}
\caption{
Number of objects that pass the selection criteria}\label{tab:data:number}
\centering
\begin{tabular}{rrrrrr}   
\hline\hline
\\[1ex] 
\multicolumn{1}{c}{$\lambda$} &
\multicolumn{1}{c}{$T_\epsilon$}  & 
\multicolumn{1}{c}{$N_{QF}$} &
\multicolumn{1}{c}{$N_\epsilon$} &
\multicolumn{1}{c}{$N_{p_V}$} &
\multicolumn{1}{c}{$N_{tax}$} \\
($\mu$m) \\
\hline

0.374 & 100 & 15592 & 828  & 828   &  642\\
0.418 & 50  & 56056 & 3460 & 3352  & 1047\\
0.462 & 30 & 59770 & 18013 & 15302 & 1268\\
0.506 & 25 & 59902 & 29025 & 23068 & 1288\\
0.550 & 30 & 60017 & 23316 & 19169 & 1278\\
0.594 & 30 & 59771 & 23181 & 19056 & 1269\\
0.638 & 25 & 58957 & 29227 & 23240 & 1258\\
0.682 & 30 & 59433 & 18920 & 16023 & 1257\\
0.726 & 25 & 59882 & 25758 & 20963 & 1280\\
0.770 & 25 & 59832 & 24576 & 20138 & 1275\\
0.814 & 25 & 59788 & 23436 & 19320 & 1274\\
0.858 & 25 & 59645 & 23026 & 19018 & 1276\\
0.902 & 25 & 59104 & 22532 & 18569 & 1268\\
0.946 & 30 & 57876 & 15728 & 13539 & 1244\\
0.990 & 60 & 55965 & 4297  & 4112  & 1086\\
1.034 & 80 & 33003 & 2257  & 2223  & 963\\
\hline\\
\end{tabular}
\vspace{1ex}

{\raggedright The condition are cumulative from left to right (see text): GDR3 quality flags ($N_{QF}$),  uncertainty threshold ($N_\epsilon$) known albedo ($N_{p_V}$),  and known taxonomy ($N_{tax}$). For each band, the central wavelength ($\lambda$) and chosen threshold ($T_\epsilon$) is also provided. \par}
\end{table}

\subsubsection{Albedo}
The albedo is crucial to discriminate between classes spectrally degenerated
\citep[e.g.,][]{1984PhDT.........3T, 2022A&A...665A..26M}.
We thus compile albedo measurements (\Cref{tab:alb_ref}) using
the \rocks\footnote{\url{https://rocks.readthedocs.io/}} Python client of the
\ssodnet service \citep{2023A&A...671A.151B}.
We found \numb{\nGaiaPv} asteroids in Gaia DR3  with albedo, among them, more than $N_{p_V}$=15,000 per band pass our threshold criteria (Table \ref{tab:data:number}). 
This sample is called ``clean sample'' hereafter. 
We also extracted other ancillary information such as
the orbital elements,
the absolute magnitude,
previous taxonomical classifications, and the family membership for each asteroid. \\

\subsubsection{Taxonomic classifications}
We used Tholen's \answer{and the recent Mahlke's schemes as a reference. Using Tholen's we take advantage of the NUV coverage of the DR3 dataset, using Mahlke's we get more recently defined taxonomies as K or L types. Furthermore, both made use of the albedo}. We therefore considered Tholen's F-class, that was not included in other taxonomic schemes \citep[e.g.,][]{2002Icar..158..146B,2009Icar..202..160D} due to the lack of NUV coverage in the dataset used to define these schemes. \answer{We kept Tholen's G-type nomenclature which in other schemes such as Bus, Bus-DeMeo's and Mahlke's was adopted as Ch-type for hydrated carbonaceous asteroids.}\\

Gaia has observed \numb{941} of the \numb{978} objects classified in \citet{1984PhDT.........3T}
and \citet{1989aste.conf..298T}. \answer{Using the} taxonomical classification of the asteroids with visible and infrared spectra in \cite{2022A&A...665A..26M}, \answer{used to} build their taxonomy, we increase the number of observations having an accurate taxonomical classification. Furthermore, we used the 25 A-type asteroids classified by \cite{2025Delbo_Atypes} as they also used VIS and IR spectra to find members of this very rare class. This increases the sample of objects classified and observed by Gaia up to \numb{1370}. From them, more than 1200 asteroids pass our quality criteria for the 12 central bands (see $N_{tax}$ in Table \ref{tab:data:number}).
There are other databases that we could use to increase the number of known objects, but we decided to keep those two as references because Tholen's is needed to follow the variations in the NUV, and Mahlke's has a wide coverage in wavelength and thus a very accurate classification. Comparison with other taxonomical works are presented in Sec.~\ref{sec:res}.\\

\subsection{Principal component analysis}\label{ssec:met:PCA}
Principal Component Analysis (PCA) reduces dimensionality by transforming the data into orthogonal directions of maximum variance. However, it cannot handle missing values requiring a complete feature matrix in the studied sample. PCA compresses information when a smaller number of Principal Components (PCs) than original variables is selected, but it also rotates the space to align axes with variations in the parameters, even if the same number of PCs as variables is selected. This is particularly useful for subsequent clustering. We applied PCA to reflectance spectra, albedo, and derived features, after standard scaling. This scaling normalizes the distribution of each variable to mean of zero and standard deviation of one preventing larger variables to have a larger variance. PCA also helps identifying noisy bands, as they often dominate the variance without carrying mineralogical information. The resulting principal component space provides a compact and structured representation suitable for clustering methods. \\
 \color{black}
 
\subsection{Gaussian mixture model}\label{ssec:met:GMM}
\color{black}
Gaussian Mixture Models (GMM) cluster data by fitting a mixture of multivariate Gaussian distributions estimated via the Expectation–Maximization algorithm. Unlike other clustering techniques, GMM performs soft clustering (i.e., each data point can belong to more than one cluster) and can capture elliptical clusters of varying sizes and orientations. However, it requires estimating many parameters, making it more efficient for uncorrelated data \citep{Bouveyron_2019}.
For this reason, we used GMM with PCA, which simplifies covariance estimation—ideally reducing it to diagonal matrices when the structure in the cluster is aligned with the axes. We then assume that each complex can be described as a mixture of normal distributions.\\
\color{black}

\subsection{Iterative clustering workflow}\label{ssec:met:iter}

\begin{figure*}
\centering
\includegraphics[width=0.98\textwidth]{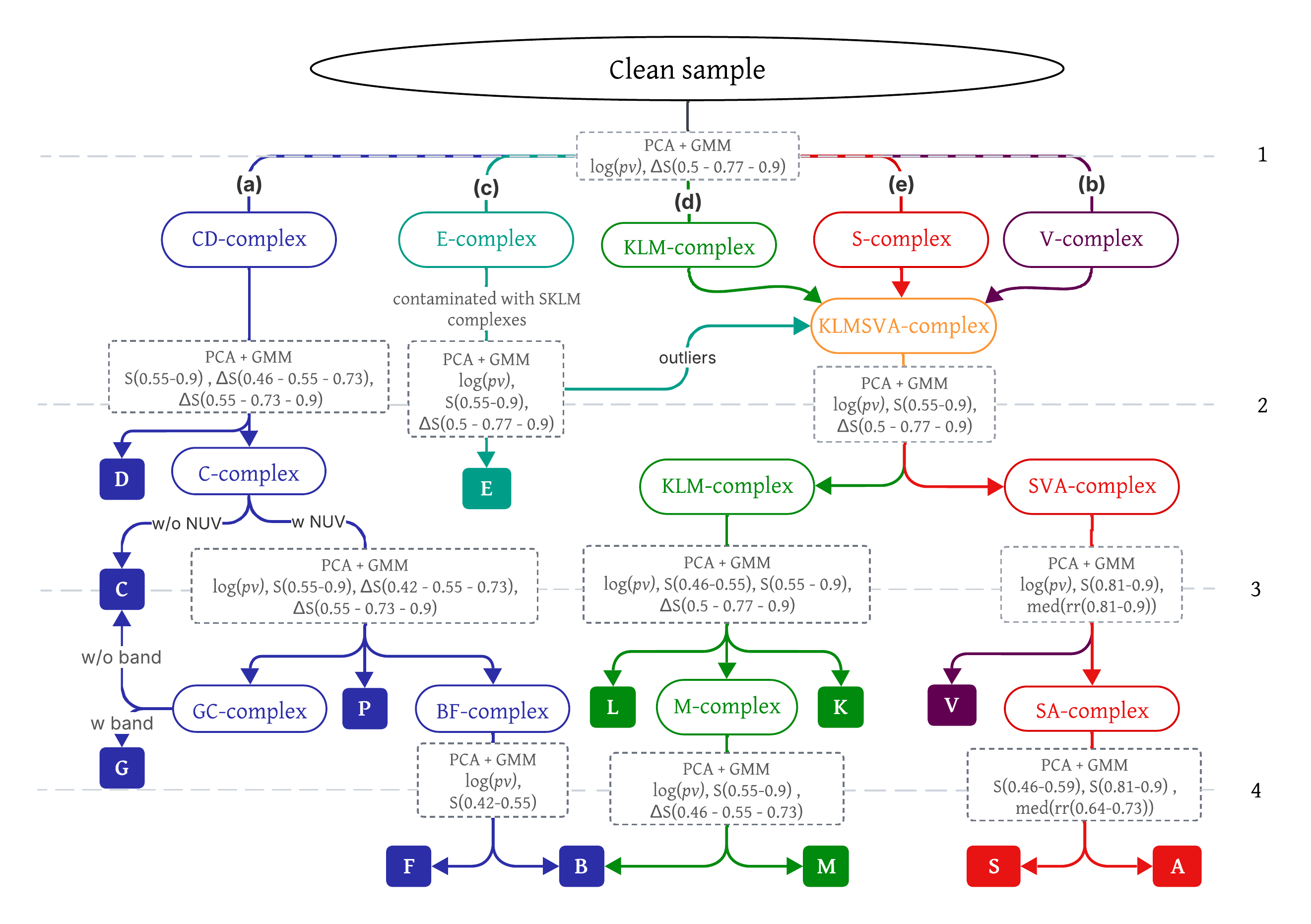}
\caption{Workflow of the taxonomic classification. Colors are related to the derived taxa and complexes
from each of the main clusters obtained in the first iteration
From left to right: the first branch (in blue) is related to the primitive asteroids with low albedo and flat spectra in the visible; the second (turquoise) is representing the high albedo Enstatite-like cluster; and the three last branches are merged into a mafic-silicate complex (green, red and purple). The deeper each branch extends, the more iterations of PCA+GMM are required to separate each taxon. We indicate the generation of the iterations with the numbered dashed horizontal lines.
}
\label{fig:met:workflow}
\end{figure*}

Previous taxonomical definitions used clustering techniques applied in a multidimensional Principal Component Space (PCS) trying to obtain all the taxa at the same time.
However, if we used this approach, applying PCA+GMM over the whole range of wavelengths, the total amount of available objects would be less than \numb{828} (see $N_{pV}$ at 0.374 in Table \ref{tab:data:number}) as PCA needs every feature available at all samples. On the other hand, if we only used the five bands with the largest number of objects (more than \numb{25,000}) we would loss the NUV information and we would not be able to find rare or specific classes such as F-class. 
Thus, we proceeded with a multi-layer classification based on PCA over computed features and an iterative clustering aiming at classifying the largest possible sample while still refining classification whenever possible, using the bands with less amount of objects at the deepest iteration possible.\\

At each layer of the classification we applied the PCA+GMM to compute spectral features known to be useful to discriminate among taxa or based on the Principal Component Bases (PCB) obtained by applying PCA over the reflectances values. The number of clusters at the different iterations may be smaller or larger than expected classes but the number is fixed to properly describe the space based on trial and errors (e.g., iteration 1 has 5 clusters sampling 3 groups). The computed features could be not available for all the objects in the previous layer, thus reducing the final classified sample at each iteration.
Each one of these clustering steps does not aim to find as many taxa as possible, but to differentiate between groups that share some of the properties, such as: flat spectra, or low albedo, or olivine/pyroxene absorption band, etc. Then, we explored each sub-cluster to search for any structure hidden in it. When the sub-clusters have no sub-structure and are not substantially different in their spectral features, albedo, and location in the Main Belt (MB), we stopped the iterative process and classify this cluster by a known taxon based on their properties.\\

The workflow from the ``clean sample'' (defined in Sec. \ref{ssec:sam}) to the final classification is summarized in Fig. \ref{fig:met:workflow} and explained in detail in the following section. The features computed for the different PCA+GMM follow this nomenclature: albedo ($p_V$); median of spectral reflectance between two wavelengths $a$ and $b$ in $\mu$m (med(${a-b})$); spectral slope between two wavelengths $a$ and $b$ in $\mu$m ($S_{a-b}$); and difference in spectral slope defined as $\Delta S_{a-b-c}=S_{a-b}-S_{b-c}$, which mainly reflects the curvature of the spectra around wavelength $b$.\\


\begin{figure*}
\import{figures}{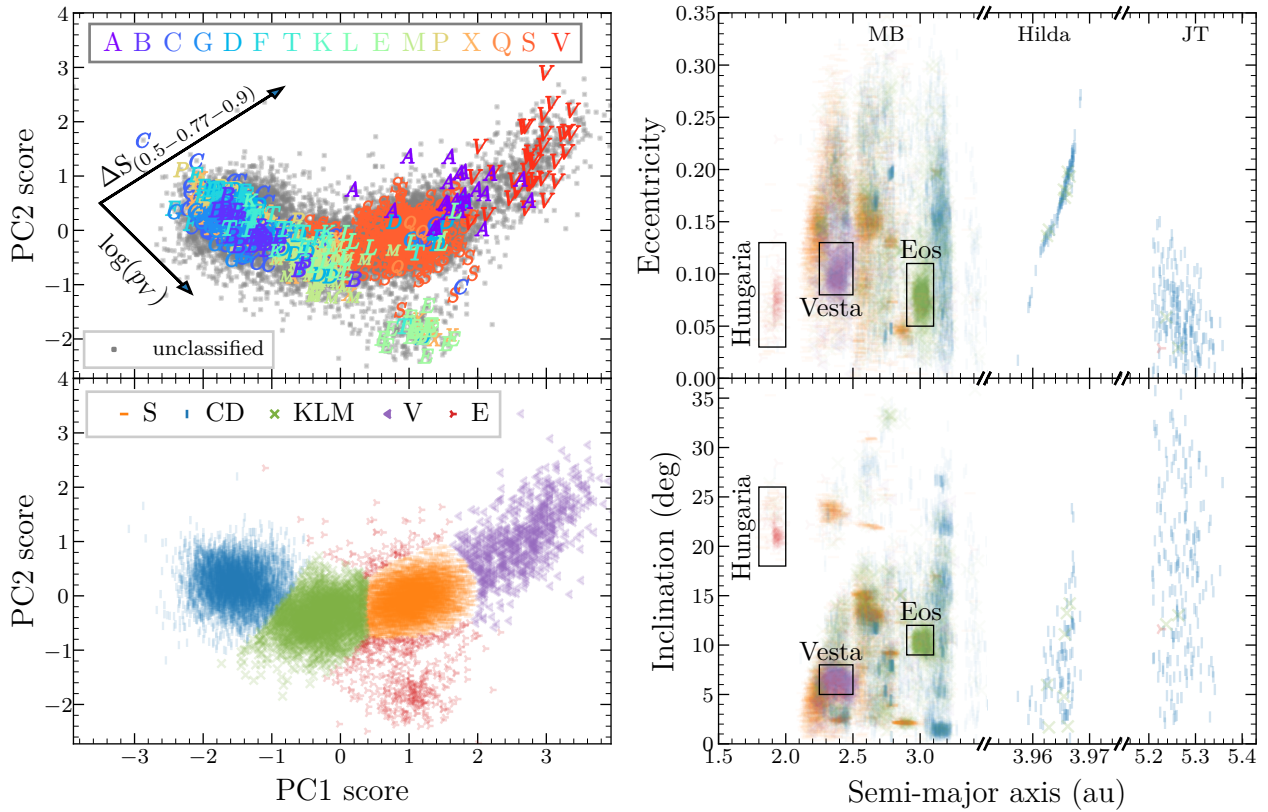}
\caption{
\textbf{Left:} First two dimensions of the PCA latent space
(using the variables $\log(p_v)$ and $\Delta S_{0.5-0.7-0.9}$ over the "clean sample").
Top panel presents previously classified objects and the bottom panel the results from the GMM clustering with the S-cluster in orange, the CD-cluster in blue, the KLM-cluster in green, the V-cluster in purple and the E-cluster in red.
\textbf{Right:} Orbital distribution of the clusters (using the same colors and symbols).
We highlight the Hungarias, Eos, and Vesta collisional families.
}
\label{fig:met:0ft} 
\end{figure*}

\section{Building the classification}\label{sec:res}


We applied a first PCA+GMM to the ``clean sample'' over two computed features: the logarithm in base 10 of the albedo \answer{\citep[$\log(p_V)$, see ][for a discussion on the use of logarithm]{2022A&A...665A..26M}} related to the presence of carbonaceous material in the surface and the difference in slope around 0.7 $\mu$m ($\Delta S_{0.5-0.7-0.9}$) related to the presence of the olivine and/or pyroxene band at 0.9 $\mu$m. Both used parameters are available for \numb{15,067} asteroids among the ``clean sample'', this is the number of sample in which is based the PCA. Even if the spectral slope over the whole wavelength range is meaningful and contains most of the variance of the sample, we decided not to use it. Indeed, using the slope calculated over the whole visible wavelength range would result in more spread clusters, making it more difficult for the GMM to identify them. Instead, the two used variables clustered together low albedo asteroids with a featureless spectra versus those with a prominent 0.9 $\mu$m band and a higher albedo.\\

We present in Fig.\,\ref{fig:met:0ft} the first iteration of the process.
Five clusters were used to sort asteroids, and they correspond to the branches in Fig.\,\ref{fig:met:workflow}.
The first cluster contains \numb{97\%} of the asteroids previously classified \answer{as} primitives (C, G, P, X, F, D, T) and is thus dubbed the primitive-complex (further described below, Sect.\ref{ssec:res:CD}).
The second cluster contains \numb{all} asteroids previously classified as E-type.
However, known E-type asteroids represent only \numb{58\%} of all objects in this cluster.
We describe the decontamination of this E-complex in Sect.~\ref{ssec:res:E}.
Finally, the three remaining clusters contain a variety of classes, from the K/L/M classes of moderate albedo
to the S/A/V classes with prominent olivine/pyroxene absorption band. We decided to merge back these three clusters together in a mafic-complex, and to separate them in a later step (Sect.~\ref{ssec:res:SKLM}).
\subsection{CD-complex}\label{ssec:res:CD}
\begin{figure*}
\centering
\import{figures}{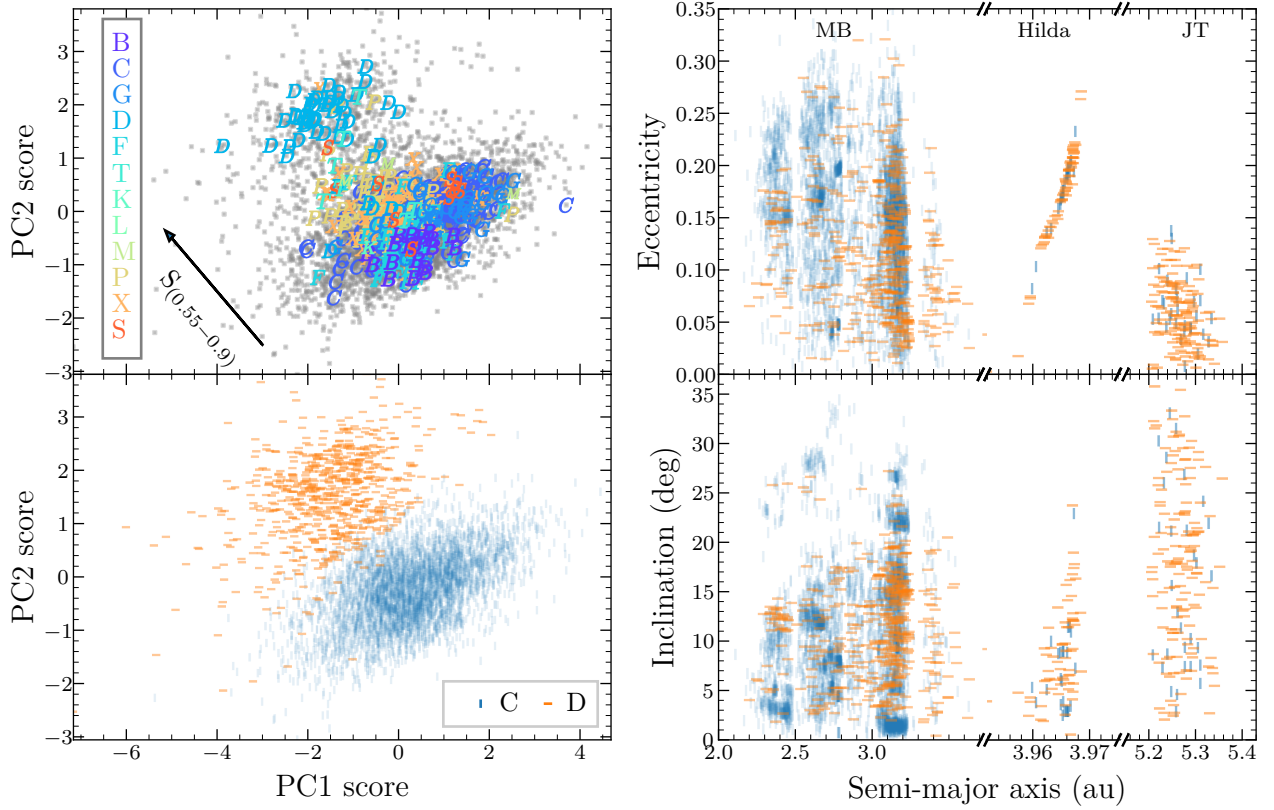}
\caption{Similar to Fig.\,\ref{fig:met:0ft} but over the primitive complex and using the $S_{0.55-0.9}$, $\Delta S_{0.46-0.55-0.73}$, and $\Delta S_{0.55-0.73-0.9}$ features.
It allows to separate the D types, in orange (dominating the Trojan space) from the other primitive taxa, in blue.}
\label{fig:met:CD}
\end{figure*}

\numb{97\%} of the previously classified objects in the CD-complex belong to the taxa C, P, D, G, F, and B. First we wanted to distinguish the D-type asteroids from the rest of the C-complex as they cluster separately in existing taxonomic schemes (e.g.,\cite{1984PhDT.........3T,2002Icar..158..106B,2009Icar..202..160D,2022A&A...665A..26M}). Those two complexes differ mainly in their slope but also in their curvature. C-complex asteroids often have a negative curvature around 0.5\,$\mu$m, while D-type asteroids are flat or have a positive curvature \citep{2024A&A...686A..76T}. Thus, we use the slope $S_{0.55-0.9}$, and the slope differences $\Delta S_{0.46-0.55-0.73}$, and  $\Delta S_{0.55-0.73-0.9}$ to disentangle D types from C types.
These features are available for 4669 asteroids among the 5257 asteroids present in the CD-complex, losing 588 asteroids (\numb{4}\% of the "clean sample").
As shown in Fig. \ref{fig:met:CD}, both clusters are well separated in this space. Applying the GMM clustering in the rotated 3-dimensional space, we found two clusters.
In the C-cluster 98\% of the classified asteroids belong to the taxa C, P, G, F, and B. On the other hand, 84\% of previously classified asteroids in the D-cluster are classified as D. In the right column of Fig. \ref{fig:met:CD}, we can see how asteroids from the C-cluster are predominant in the MB, while the D-cluster is predominant in Hildas and Trojans, as is expected for C-complex asteroids and D-type, respectively \citep{2013Icar..226..723D}.\\

\begin{figure}
\centering
\import{figures}{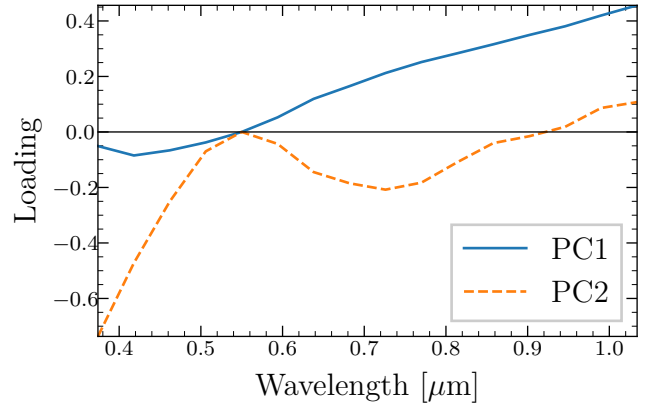}
\caption{Loadings of the first two Principal Components from the base resulted after the PCA of the reflectance spectra in the CD-complex. Both features discussed in the text, 0.7 $\mu$m band and UV absorption appear together with the same sign in PC2, in orange.}\label{fig:met:C_NUV}
\end{figure}

As no substructure in found in the D-cluster, we defined it as the D taxon. For the C-complex we found a PCB with a first principal component related to the slope, and a second one, different from previous studies, related to the coexistence and correlation of the NUV absorption and the 0.7 $\mu$m band (see Fig. \ref{fig:met:C_NUV}). Third and fourth PCB are related to noise in bands at 0.506 $\mu$m and 0.594 $\mu$m respectively and are not showed in the figure. In the space generated by the first two principal components, we were able to find known F, P, G, and C-types differently distributed in the PCS. However, PCA cannot handle missing values and only 391 C-complex objects in Gaia DR3 have spectra in which all bands meet our quality thresholds. In addition, most of them are already classified. To increase the number of asteroids that PCA could handle, and thus we could classify, we computed a set of variables that allowed us to characterize the features found in the PCB. We use the set of four features: log$_{10}$($p_v$), $S_{0.55-0.9}$, $\Delta S_{0.42-0.55-0.73}$, $\Delta S_{0.55-0.73-0.9}$. Both differences of slopes account for the change in slope at 0.55 $\mu$m and 0.73 $\mu$m, i.e., the UV absorption and the 0.7 $\mu$m band. Those features are available for 1121 asteroids among the 3967 asteroids in the C-complex (\numb{28}\%); the band at 0.42 $\mu$m being the most limiting. For the asteroids in the C-complex for which the 0.42 $\mu$m band has not passed our selection criteria, we assign them to the C-type, as we do not have enough information to discriminate between the different taxa found in the C-complex.\\

\begin{figure*}
\centering
\import{figures}{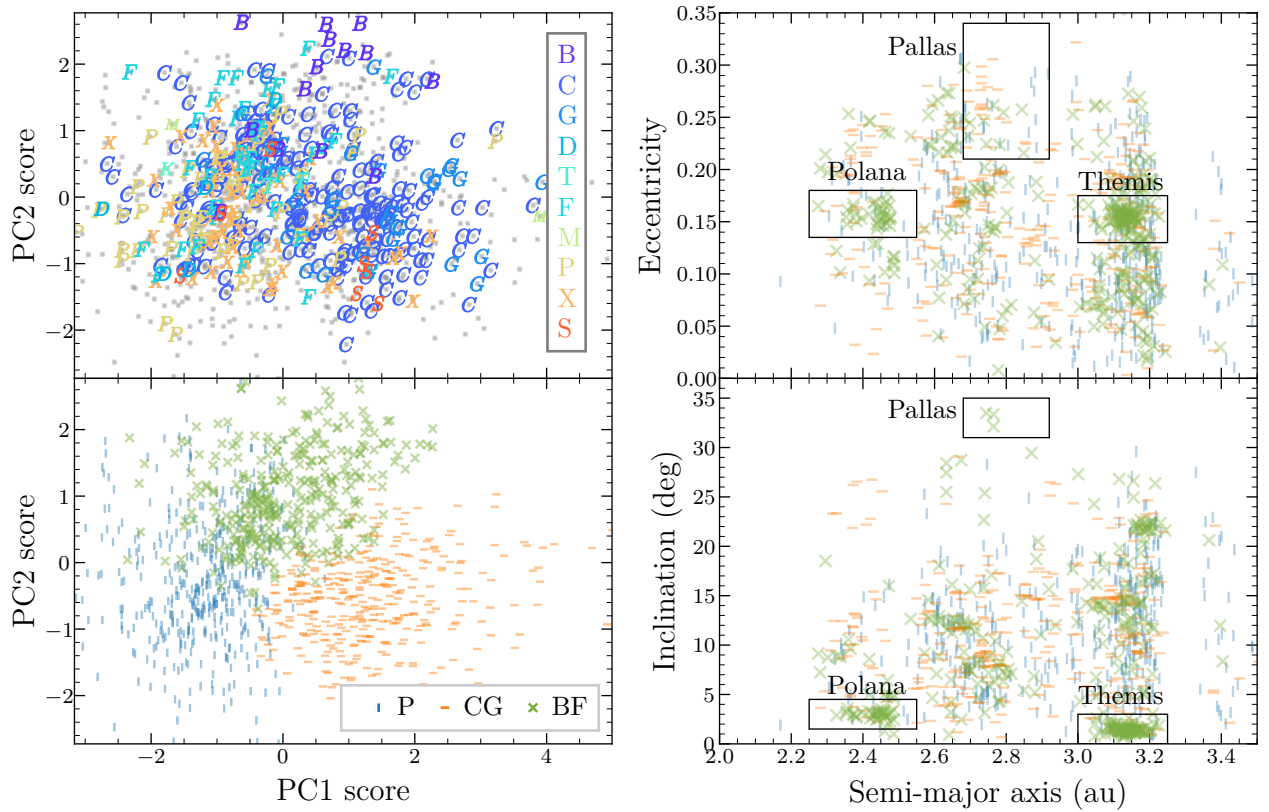}
\caption{Similar to Fig.\,\ref{fig:met:0ft} but over the C-complex from the second iteration and using four features: log$_{10}$($p_v$), $S_{0.55-0.9}$, $\Delta S_{0.42-0.55-0.73}$, and $\Delta S_{0.55-0.73-0.9}$. This iteration efficiently splits asteroids into three complexes: green, orange and blue, associated with BF, CG and P classes respectively.
We highlight the F-type Polana family and the B-type Themis and Pallas families on the right, revealing a dwarf of objects among the latter family (see text).}
 \label{fig:met:C}
\end{figure*}

After applying a PCA to reduce the four original dimensions to three,
we find three clusters inside the C-complex (Fig.~\ref{fig:met:C}). These clusters correspond to P, C-G, and B-F taxa.
In the C-G group we found that most of the asteroids were previously classified as C type, but 80\% of the asteroids previously classified as G or Ch are in this cluster. To discriminate among G (Ch) and C, we defined another feature: we made use of the band depth at 0.7 $\mu$m, computed by \cite{2024A&A...686A..76T}. This band is the main feature of the visible spectra of G or Ch-type asteroids \citep{1984PhDT.........3T, 2002Icar..158..106B}. The objects in the cluster having a band depth larger than \numb{1\%} were classified as G-types. The median spectrum and albedo of taxon G are displayed in Fig. \ref{fig:res:med_esp}. The rest of the objects are classified as C-types and incorporated into the taxon C. Finally, the remaining cluster contains a mixture of objects previously classified as P, X, and C-types. According to the median spectra of this cluster (pannel P in Fig. \ref{fig:res:med_esp}), they are redder than the other clusters (pannels G, C, B or F in Fig. \ref{fig:res:med_esp}) and have a very shallow to none absorption below 0.55 $\mu$m. When applying a PCA and GMM over the objects in this cluster, we were not able to find any sub-structure, thus, we decided to define the taxon P as members of this cluster.
There are 53 asteroids previously classified in this cluster by \cite{2022A&A...665A..26M} with both visible and IR data. From them 12 were previously classified as C, four as Ch (G-type in our scheme) and 33 as P-type. Even if we assume some false positives in this group, we suggest that use of NUV allowed us to distinguish efficiently the taxa because is tracing the presence of the UV absorption (indicative of C or G types) versus the lack of absorption (indicative of P types).\\

\begin{figure*}
\centering
\import{figures}{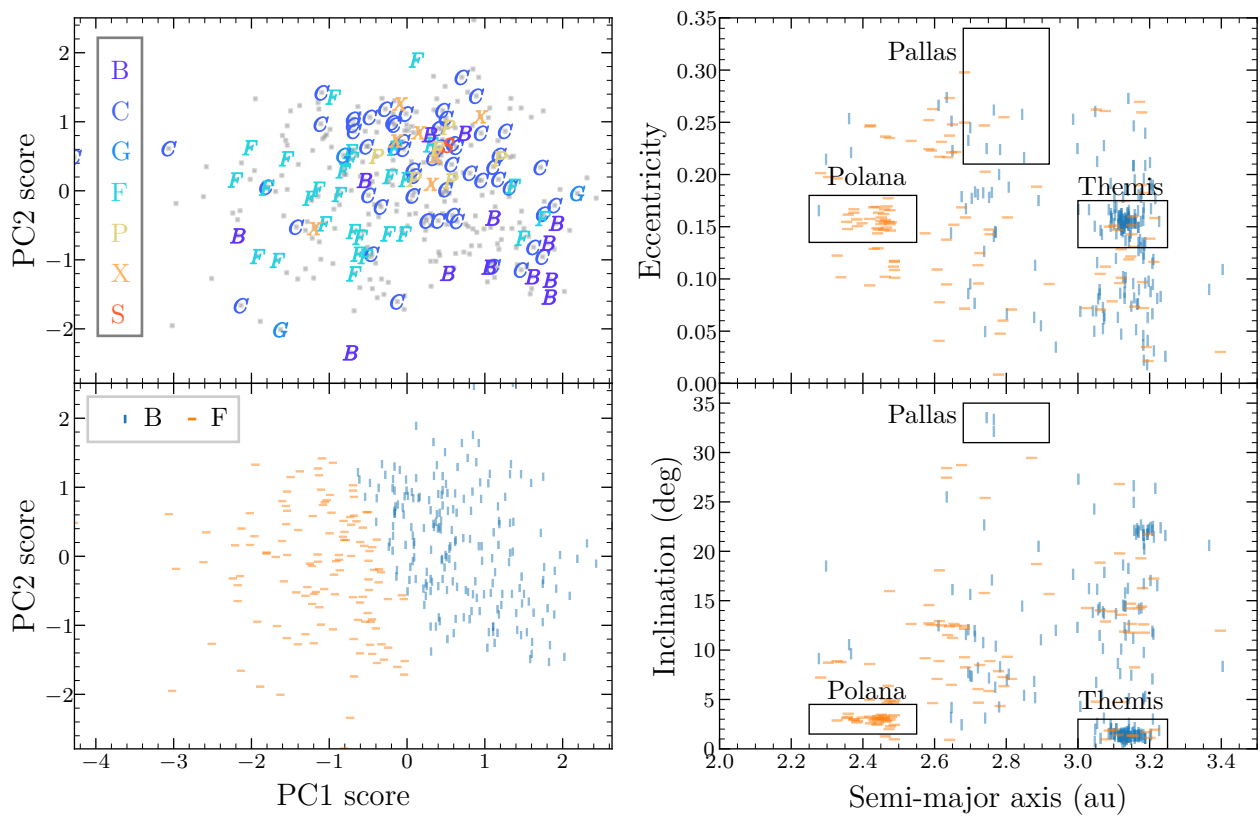}
\caption{Similar to Fig.\,\ref{fig:met:0ft} for the BF-complex and with the  features: $S_{0.55-0.9}$ and log$_{10}$($p_v$), showing the divergence between the B and F taxa (in blue and orange, respectively). The result is visible from the concentrations of different taxa in the Polana, Themis and Pallas families (right).
}
\label{fig:res:BF}
\end{figure*}

As a final step for the study of primitive asteroids, we distinguished B-types from F-types. They differ
in their slopes between 0.4 and 0.5 $\mu$m, the first having an absorption through the NUV beginning at 0.5 $\mu$m, the latter remaining flat until 0.4 $\mu$m, where their absorption begins \citep{1984PhDT.........3T, 2024A&A...686A..76T}. Furthermore, B-types asteroids are known to have a higher albedo than other carbonaceous asteroids\footnote{hence the ``B'' for \textbf{B}right cornered by \citet{1984PhDT.........3T}}.
Thus, we use the features $S_{0.55-0.9}$ and log$_{10}$($p_v$) to distinguish those two taxa, available for all asteroids present in the BF-complex. Fig. \ref{fig:res:BF} presents the clustering and orbital distribution of the asteroids in the BF-complex. While the separation between the two clusters is not entirely obvious, the GMM finds two clusters that correspond to two different asteroid families in the MB.
The F-cluster is populating the Polana-Eulalia region, a family known to be mainly populated by F-type objects \citep{2022A&A...664A.107T}. The B-cluster is concentrated in the Themis family, known to be rich in B-type asteroids \citep{2022A&A...664A.107T}. The albedo distribution and median spectra of asteroids on F and B-clusters (Fig. \ref{fig:res:med_esp}) are coherent with the literature for F and B-types. Therefore, we defined the B and F-clusters as the B and F-taxon, respectively.\\ 

Something to point here is the low number of objects in the Pallas family belonging to the B-cluster, when Pallas is known to be a B-type family \citep{2022A&A...664A.107T}. We noticed that Pallas family members had been misclassified as members of the KLM-complex after the first iteration of the classification, as can be seen in the region around (a$\approx$2.6 au, i$\approx$33 (deg)) at Fig. \ref{fig:met:0ft}. \answer{This confusion is due to the use of a difference of slope in the first iteration, as B-type objects can be bright and without absorption around 0.7 microns, as the M-type, thus sharing the PC space with them. At furthest steps we are using the value of the slope (not the difference) allowing the algorithm to differentiate the brightes B-types from the M-types (see Sect. \ref{ssec:res:SKLM}).} Thus we were able to reinject missclassified B-types into the taxon B, classifying Pallas family members as B-types consistently with the literature.

\subsection{E-complex}\label{ssec:res:E}


While dominated by E types, the E complex is contaminated by objects similar to those from the mafic complex (K, L, S...).
We identify these outliers by applying a new PCA + GMM clustering over three features:
log$_{10}$($p_v$),
$\Delta S_{0.5-0.7-0.9}$,
to which we added the slope in the visible ($S_{0.55-0.9}$)
because S- and E-type have different visible slopes \citep{1984PhDT.........3T, 2022A&A...665A..26M}.
Those features are available for all the asteroids in the E-complex.
This iteration allows to perfectly isolate all known E-types from contaminants (Fig. \ref{fig:met:E}).
Asteroids belonging to this cluster are mainly populating the Hungaria region, known to be rich in E-type asteroids \citep{2019Icar..322..227L}. 
The other objects are re-injected in the mafic complex obtained from the first iteration, to be further classified at subsequent iterations.

\subsection{Mafic-complex}\label{ssec:res:SKLM}

This cluster is the result of the merging of the non-primitive non-E-Type complexes (i.e., KLM, V and S-complexes) of the first clustering iteration, with the outliers obtained from the second iteration on the E-complex. We first split KLM- from SVA- related material using the 
log$_{10}$($p_v$),
$\Delta S_{0.5-0.7-0.9}$),
and $S_{0.55-0.9}$ (visible spectral slope) features,
available for all asteroids in this cluster
(Fig.\,\ref{fig:met:SKLM})

\begin{figure*}
\centering
\import{figures}{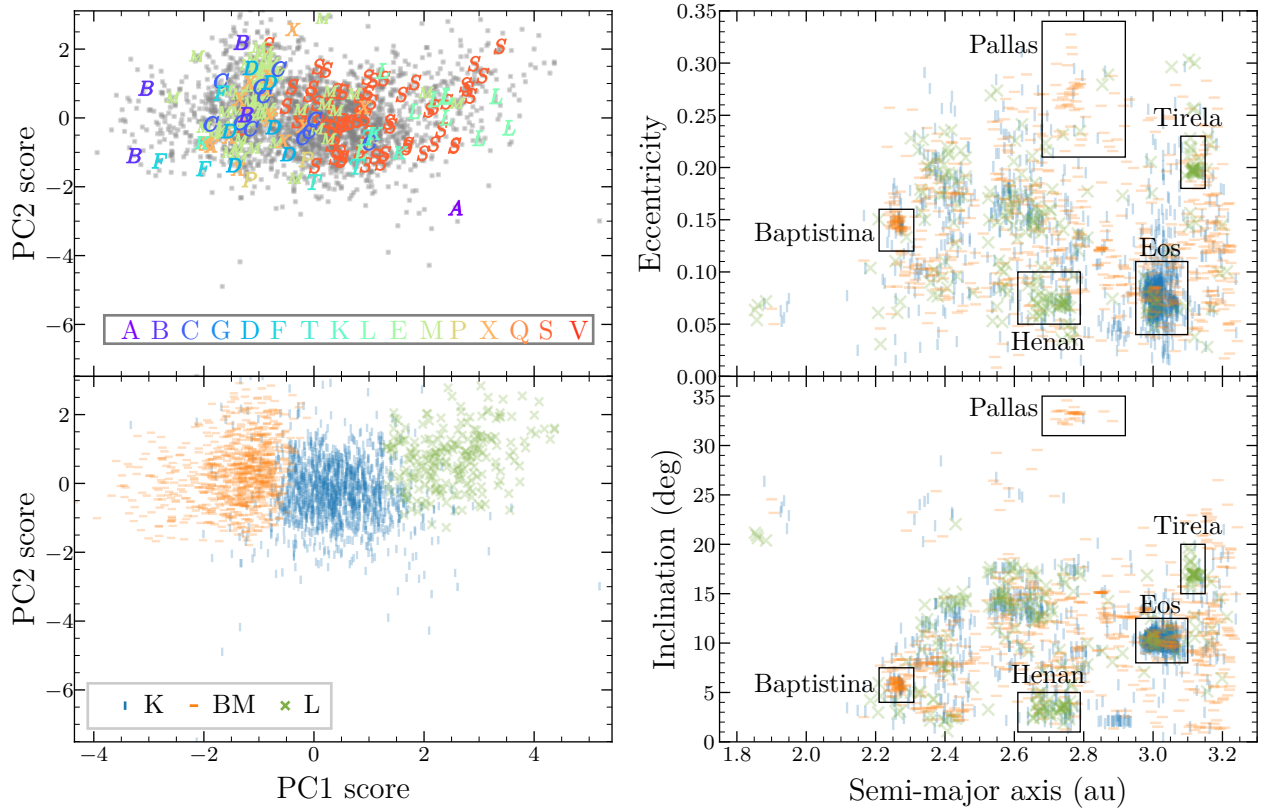}
\caption{Similar to Fig.\,\ref{fig:met:0ft} but for the KLM-complex and using four features: log$_{10}$($p_v$), $\Delta S_{0.5-0.7-0.9}$, $S_{0.5-0.9}$, and $S_{0.46-0.55}$.
The main clusters are related to K taxon (blue) and the Eos family, the L taxon (green) with 
the Tirela and Henan families, 
a mixture of B and M types (orange, see the Baptistina and Pallas families)}
\label{fig:met:KLM}
\end{figure*}

Among the KLM complex, the 
most descriptive features are the
log$_{10}$($p_v$), $\Delta S_{0.5-0.7-0.9}$, $S_{0.5-0.9}$ used in previous step
together with the spectral slope between 0.46 and 0.55\,$\mu$m ($S_{0.46-0.55}$).

Those features are available for \numb{2342} asteroids among the \numb{2515} asteroids in the KLM-complex (\numb{93}\%), loosing 173 asteroids (\numb{1}\% of the "clean sample").
As visible in Fig. \ref{fig:met:KLM}, we efficiently identify 
K- and L-type asteroids.
First about \numb{80\%} of the asteroids classified in the Eos family
\citep[a well-known K-type family,][]{1989Icar...78..426B, 2000Icar..145....4Z}
belong to a single cluster and they do not have much prevalence outside this family.
We thus defined the K taxon from this cluster.
Second, many objects belonging to L-type families Tirela and Henan \citep{2024A&A...688A.221B}
lay in another cluster, which thus represent the L taxon.
Finally, the objects in the third and last cluster are abundant 
in some families with M-type parent bodies
\citep[e.g., Baptistina][]{2013A&A...551A.117B, 2022A&A...665A..26M} and 
among the Pallas family \citep[made of B types,][]{2016A&A...591A..14A}.
The B types with their high albedo (compared with other primitive types) and mainly featureless spectra
are somewhat reminiscent of M (for metallic) types, and thus contaminate this cluster
(while being made of entirely different mineralogies).

We thus used other spectral features to differentiate B- and M-types:
the visible slope ($S_{0.5-0.9}$, M- being generally redder than B-types),
the difference in slope at 0.55 $\mu$m ($\Delta S_{0.46-0.55-0.73}$,  B-types show an decrease in reflectance in  the NUV related to iron-rich phyllosilicates, not present in M-type asteroid spectra),
and the albedo (log$_{10}$($p_v$)).
These features are available for all asteroids in the BM-complex.
As illustrated in in Fig.\,\ref{fig:met:BM},
the B types cluster in both the PC space and orbital elements (Pallas and Themis families), and we
grouped them with the taxon B (obtained in the last step of Sect.\,\ref{ssec:res:CD}).
Conversely, we defined the M taxon from the other cluster, populating other regions of the MB, such as the Baptistina family.\\



\begin{figure*}
\centering
\import{figures}{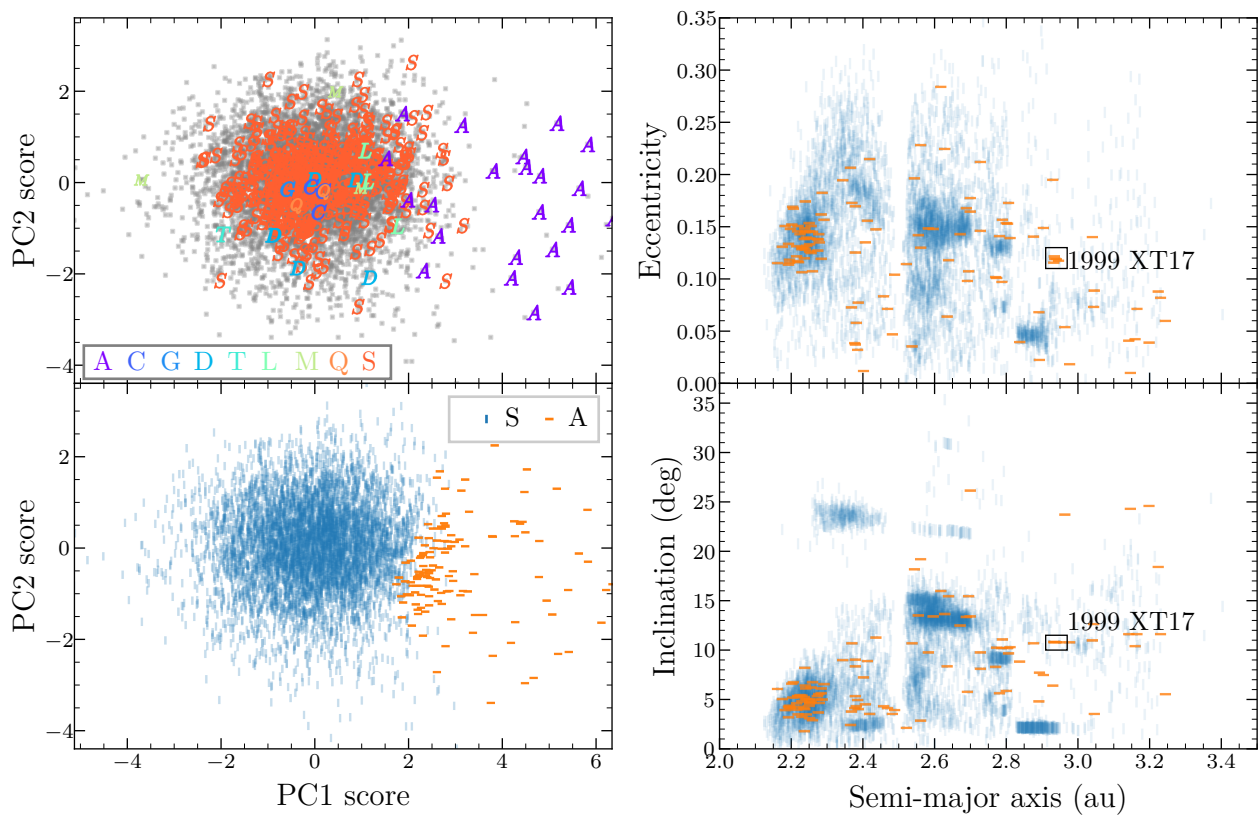}
\caption{Similar to Fig.\,\ref{fig:met:0ft} but for the SA-complex with three features $S_{0.46-0.59}$, $S_{0.81-0.9}$ and med(${0.64-0.73}$) to 
separate the A taxon (orange) from the S taxon (blue).
We highlight in the right panels, the first collisional family known to be related to A-type asteroids: 1999\_XT17.
%
}
\label{fig:met:SA}
\end{figure*}

The other sub-group raising from the second iteration is the SVA cluster
(Fig.\,\ref{fig:met:SKLM}) from which we first identified the V types, using the
features log$_{10}$($p_v$), $S_{0.81-0.9}$ and med(${0.81-0.9}$), available 
for all asteroids in the SVA complex.
The results are shown in Fig. \ref{fig:met:SV}. 
We defined the V types from a single cluster, which densely populates the Vesta family.
The other cluster encompasses both S- and the A-type asteroids that we split using
the slope $S_{0.46-0.59}$ \citep[very reliable to identify A types in Gaia DR3 following ][]{2024PhDT.........G},
$S_{0.81-0.9}$ and med(${0.64-0.73}$), describing the beginning of the 0.9 $\mu$m band).
Those features are available for 6043 asteroids among the 6307 asteroids in the SA-cluster (\numb{96}\%).
We defined the A taxon from the A-cluster 
(Fig.\,\ref{fig:met:SA}) that contains \numb{86\%} of previously known A-types.
Comparing the most-recent focused work on A-types \citep{2025Delbo_Atypes}, this fraction raises to 13 out of their 14 A types. Furthermore, this cluster contains the first family rich in A-type asteroids found by \citet{2024A&AGalinier_A_family}. 
Finally, the other cluster contains \numb{93\%} of objects previously classified S. As it does not present any substructure, this cluster defines the S taxon.\\

We have to remember that after every taxon definition, we have checked their possible substructures by doing a PCA over the reflectance spectra, and no more substructures related to known taxa are found.\\

\color{black}

\section{Results}\label{sec:res2}
\subsection{Classification outcome}\label{ssec:res:class}

\begin{figure*}
\centering
\import{figures}{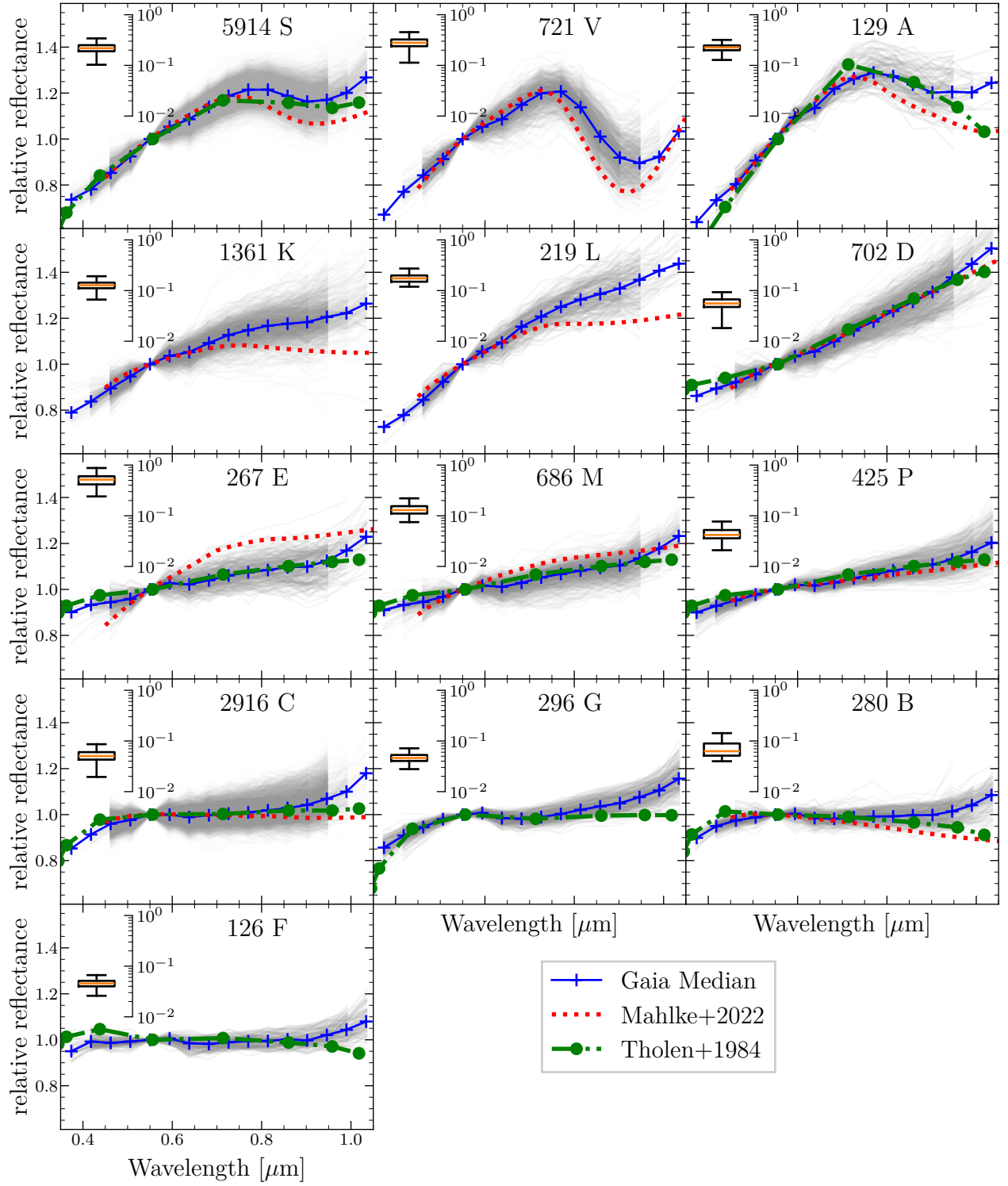}
\caption{Median spectra (blue solid line) Gaia reflectance spectra computed from individual spectra (grey) for each of the 13 obtained taxa. For comparison, we added the Tholen and Mahlke reference spectra for the common taxa between our study and theirs (in dashdoted green and dotted red lines). We also included a boxplot of albedo for each taxon. The total number of objects classified is in the title of each subplot.
}\label{fig:res:med_esp}
\end{figure*}

\color{black}
After the cleaning processes explained in section \ref{ssec:sam} and the methods explained in Sec. \ref{sec:met} we classified successfully \numb{\nClassif} asteroids. To test the goodness of this classification, we compared the median spectra of our classified taxa with previous reference spectra in Fig. \ref{fig:res:med_esp}. This comparison showed the goodness of our threshold cuts as the discrepancy in the NUV between the Gaia DR3 and ground-based spectra is not noticeable. However, the reddening of the Gaia DR3 spectra in the wavelengths between 0.7 and 0.9 $\mu$m \citep{2023MT........B, 2023A&A...671A..40G, 2023Oszkiewicz_basaltic_Gaia, 2026MNRAS.545f2052M} is clear also \answer{in some taxonomies} among our results. This reddening is present in every taxon but D, E, M, P \answer{and F}, the spectral types corresponding to the most featureless spectra. These differences are useful to highlight that the final median spectra from Gaia DR3 taxa should not be directly compared or classified without caution with reference spectra from other sources. However, we can see how main features trends among taxa and qualitative description of the characteristics of every taxon are still in agreement with previous taxonomies showing
a deeped 0.9 $\mu$m band
LK$\rightarrow$A$\rightarrow$S$\rightarrow$V,
an increasing visible slope for primitives BF$\rightarrow$CG$\rightarrow$P$\rightarrow$D,
an increasing albedo P$\rightarrow$M$\rightarrow$E with a similar spectral shape,
or an increasing the prevalence of the 0.7 $\mu$m band from C$\rightarrow$G.\\

\begin{figure*}
\centering
\includegraphics{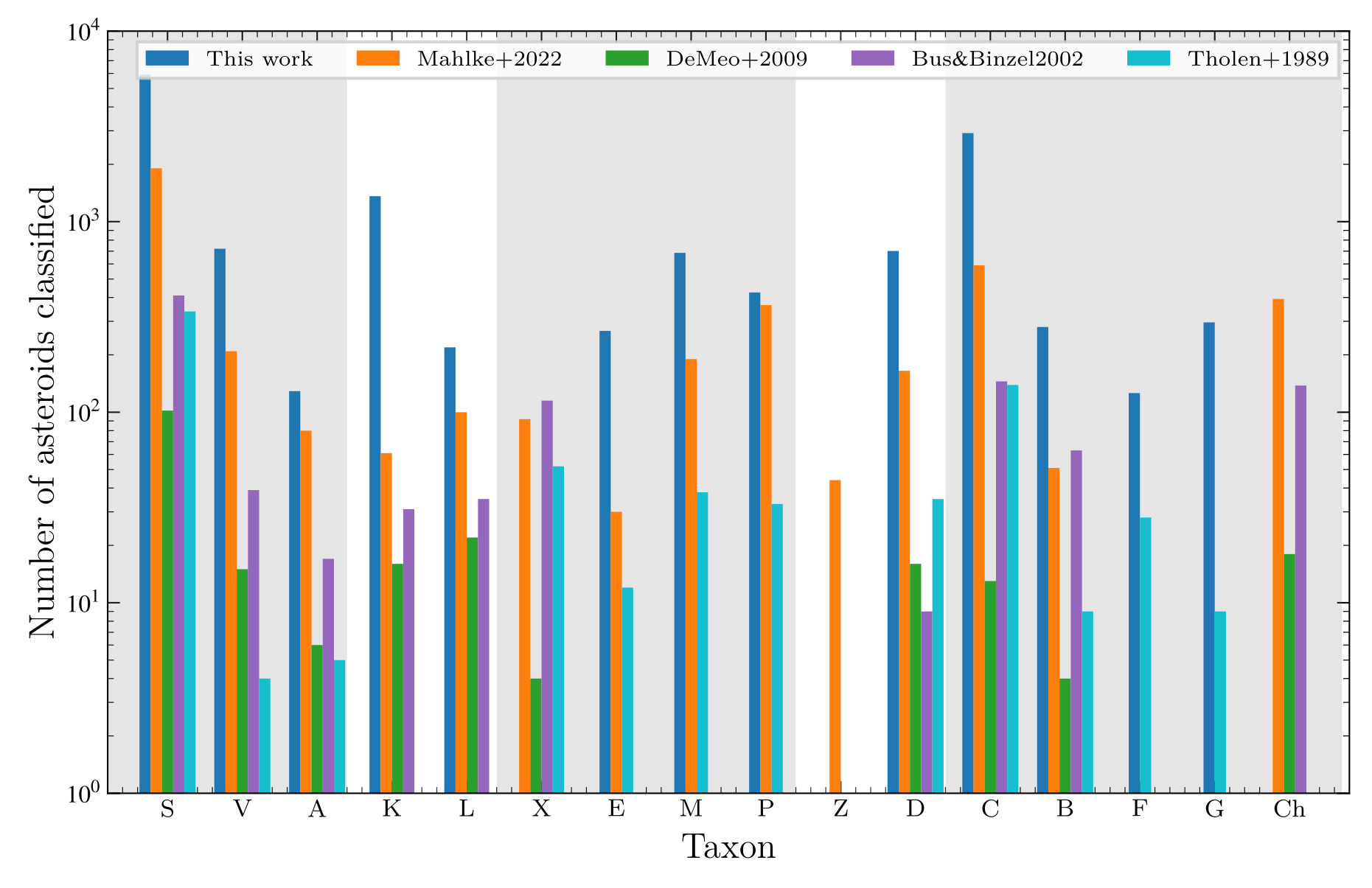}
\caption{Histogram showing number of asteroids classified per taxon comparing this work with \cite{2022A&A...665A..26M} in orange, \cite{2009Icar..202..160D} in green, \cite{2002Icar..158..106B} in purple, and \cite{1984PhDT.........3T, 1989aste.conf..298T} in cyan.}
\label{fig:res:comp}
\end{figure*}

Compared to the most recent taxonomical classifications that used spectral data \citep{2022A&A...665A..26M}, we increased the sample of classified objects by a factor of \numb{seven}. We find fewer objects for the G taxon, when compared to Ch-types from \cite{2022A&A...665A..26M}. This is due to our need to use the Gaia band at 0.418 $\mu$m to succesfully classify them as G types. This band is not present for most of the asteroids, see column $N_{pv}$ at row $\lambda$=0.418 $\mu$m in Table~\ref{tab:data:number}. Our largest increase is for K-types, where we classified \numb{28} times more objects than in \cite{2022A&A...665A..26M}. A detailed comparison with previous spectral taxonomical classifications, such as \cite{1984PhDT.........3T, 2002Icar..158..146B, 2009Icar..202..160D, 2022A&A...665A..26M} is presented in Fig. \ref{fig:res:comp}. 

The increment of the present taxonomical classification per taxon compared with previous taxonomical classifications is significant.
We plotted closer taxa with similar spectra or features (S-V-A, K-L, X-E-M-P, D-Z, C-B-F-G-Ch).
We notice that we did not classify any X type \citep{2002Icar..158..106B,2009Icar..202..160D} because the albedo is available for all asteroids allowing to distinguish between E, M, and P.
We did not find any structure in the D-cluster, so we were not able to find the taxon Z defined in \cite{2022A&A...665A..26M}.
Finally, as we used the NUV information and the nomenclature from \cite{1984PhDT.........3T}, we were able to distinguish F-types from the C-complex, that are not present in any other taxonomic scheme apart from Tholen's. Following \cite{1984PhDT.........3T}, we also chose to define a G-class instead of the Ch-class of later taxonomies \citep{2002Icar..158..106B,2009Icar..202..160D,2022A&A...665A..26M} to encompass the group of hydrated carbonaceous asteroids with their characteristic absorption band at 0.7 $\mu$m. Some numbers to highlight in this graph is the large number of K-type asteroids that we have found, most of them coming from the Eos family. We suspect that this is due to the non-targeted pointing of Gaia. Previous datasets were compiled from dedicated observations of individual objects.
However, Gaia provides a coverage of the sky without a bias due to target selection, thus, it observed more objects from the large families such as Eos or Vesta, increasing the amount of their own taxonomical types. On the other hand, the large number of new C-types, is due to our classification criteria, as we put in this taxon those asteroids in the C-complex that had not passed the cleaning selection for the spectral point at 0.42 $\mu$m and thus can not be differentiated among F, B, C, G or P taxa. Looking at Fig. \ref{fig:res:med_esp}, it can be seen that the median spectra of C-types follows the behaviour previously described for this taxon (flat, featureless spectra) \answer{but the dispersion of the individual spectra is higher in this taxon as most of them come from C-complex spectra whose 0.418 $\mu$m point did not pass the cleaning criteria, having a smaller SNR.}\\


\color{black}
\subsection{Taxonomical distribution in dynamical populations}\label{ssec:res:dist}

\begin{figure*}
\centering
\import{figures}{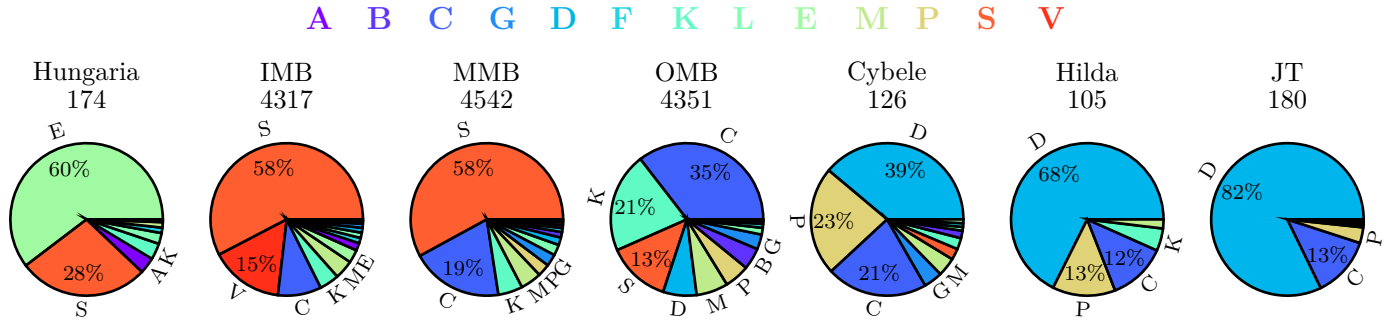}
\caption{Distribution of taxa per dynamical population. The color code follows the same pattern as the previous graphs. Taxa more prevalent than \numb{3\%} are labeled and their percentage in the population shown if larger than \numb{10\%}. The total number of asteroids classified in the present study for each population is also reported.}
\label{fig:res:tax_dist}
\end{figure*}

One direct result from the taxonomical classification comes from inspecting the distribution of taxa along the dynamical populations. We compared the results from this analysis with \cite{2013Icar..226..723D} as a test of the goodness of our classification. \citet{2013Icar..226..723D} used SDSS photometry reaching a larger sample size but a lower spectral resolution. In Fig \ref{fig:res:tax_dist}, we can see pie charts with the percentage of each taxonomy in the different dynamical regions, following Fig. 4 in \citet{2014Natur.505..629D}. We can see \numb{61\%}  of Hungarian asteroids are E-types, the \numb{33\%} are S-type asteroids. However E-types are hardly present in the rest of the MB. S-type asteroids dominate the Inner and the Middle Main Belt (IMB and MMB) with a \numb{60\%} in both cases. The presence of primitive asteroids (C, P, D, B, F, G) keeps growing from \numb{11\%} in the IMB to \numb{55\%} in the Outer Main Belt (OMB) with a \numb{25\%} in the MMB. This is in agreement with our current understanding of MB formation and evolution \citep{2015aste.book..493M, 2015ApJ...806..143V, 2009Natur.460..364L}.
If we continue looking further than the MB the primitive objects dominate over \numb{90\%}, with the D as the more usual taxon. \answer{D-type represents} \numb{39\%} among the Cybeles, \numb{68\%} among the Hildas, and \numb{83\%} among the Jupiter Trojans (JTs), where the primitive objects dominate with \numb{99.4\%} of the sample. The number of D-types is in good agreement with the recent results from \cite{2025Noemi_Dtypes}, and \cite{2026Noemi_Dtypes} for the IMB, the MMB, the Hildas, and the JT. They found, using Gaia DR3 data, a 0.6\% of D and Z-types in the IMB, a 1\% in the MMB, 69\% of D and Z-types in Hilda, and an 80\% in the JTs. However they found a slightly larger amount of D and Z-types in the Cybele region (a 45\% in their work versus 39\% in ours). On the other hand, we found a slightly larger amount in the OMB (7\% in this work versus 2.3\% in \cite{2026Noemi_Dtypes}). Those small discrepancies could be explained as the original sample in both works was not the same, as the threshold used in the noise of the bands is not the same.\\


Another interesting point is the effect of the large families Vesta and Eos, which populate the IMB and the OMB with V and K types, respectively, making those taxa the second most common in their respective populations with a \numb{15\%} and a \numb{21\%} of prevalence.\\

We have to highlight here that, as any other magnitude-limited surveys, we have an observational bias against small, dark, and far objects that could be affecting the real distribution of taxa. High albedo taxa are observed further and/or smaller in their real distribution than low albedo taxa.\\

\section{Conclusions}\label{sec:con}

In conclusion, we have classified the reflectance spectra of \numb{\nClassif} Gaia DR3 asteroids. However, this data release has some artifacts that requier caution when performing direct comparison with reference spectra obtained in previous classifications. Thus, to classify Gaia DR3 we needed to follow a process of clustering as if we were trying to build a new taxonomy, and then relate clusters to known taxa. To achieve our objective, we first cleaned the sample based on the quality of the spectral bands, evading some of the artifacts related to the signal to noise ratio. Then, we carried out an iterative process of clustering that allowed us to find 13 taxa related to known: A, B, C, D, E, F, G, K, L, M, P, S, V. The total amount of classified objects is \numb{\nClassif} asteroids becoming the largest systematic classification using spectral information.  We must emphasize the relevance of the presence of NUV information among the Gaia data allowing us to distinguishing between the taxa in the C-complex (i.e., B, F, G taxa). As a test of the classification we inspected the distribution of taxa in the dynamical populations and collisional families of the Solar System. Their distribution is in agreement with previous results, reinforcing our trust in our classification method. When comparing our classified objects with previous spectral classifications, we noticed a reddening in the spectra between 0.7 and 0.9 $\mu$m of Gaia DR3 spectra, already noticed by other authors, reminding us that neither the obtained median spectra should be used to classify other objects outside the Gaia DR3 catalog, nor previous templates should be used to classify Gaia DR3 spectra without accounting for this reddening. \\

This classification offers a unique opportunity to explore the heterogeneity in the collisional families and refine their definition based on the taxonomical classification as well as inspecting the spatial distribution of material in the Solar System.\\

\section{Data availability}
Table with the classification outcome are only available in electronic form at the CDS via anonymous ftp to cdsarc.u-strasbg.fr (130.79.128.5) or via http://cdsweb.u-strasbg.fr/cgi-bin/qcat?J/A+A/

\begin{acknowledgements}
Thanks to Noémie el Bez Sebastien for checking the goodness of our D-type asteroids, and Roberto Balossi for sharing his knowledge about L-type families. This work has made use of data from the European Space Agency (ESA) mission Gaia (\url{https://www.cosmos.esa.int/gaia}), processed by the Gaia Data Processing and Analysis Consortium (DPAC, \url{https://www.cosmos.esa.int/web/gaia/dpac/consortium}).
F.T-R thanks CNES for funding my postdoctoral program.
This work was supported by the Programme National de Planétologie (PNP) of
CNRS-INSU co-funded by CNES.
This work used Virtual Observatory tools
SSODNet \citep{2023A&A...671A.151B} and
TOPCAT \citep{2005ASPC..347...29T}.
Thanks to the developers and maintainers.
\end{acknowledgements}

\bibliographystyle{aa}
\bibliography{refs}

@INPROCEEDINGS{1973BAAS....5..388Z,
       author = {{Zellner}, B.},
        title = "{Polarimetric Albedos of Asteroids.}",
    booktitle = {Bulletin of the American Astronomical Society},
         year = 1973,
       volume = {5},
        month = sep,
        pages = {388},
       adsurl = {https://ui.adsabs.harvard.edu/abs/1973BAAS....5..388Z}
}

@ARTICLE{1975Icar...25..104C,
       author = {{Chapman}, C.~R. and {Morrison}, D. and {Zellner}, B.},
        title = "{Surface Properties of Asteroids: A Synthesis of Polarimetry, Radiometry, and Spectrophotometry}",
      journal = {Icarus},
         year = 1975,
        month = may,
       volume = {25},
       number = {1},
        pages = {104-130},
          doi = {10.1016/0019-1035(75)90191-8},
       adsurl = {https://ui.adsabs.harvard.edu/abs/1975Icar...25..104C}
}

@ARTICLE{1985Icar...61..355Z,
       author = {{Zellner}, B. and {Tholen}, D.~J. and {Tedesco}, E.~F.},
        title = "{The eight-color asteroid survey: Results for 589 minor planets}",
      journal = {Icarus},
         year = 1985,
        month = mar,
       volume = {61},
       number = {3},
        pages = {355-416},
          doi = {10.1016/0019-1035(85)90133-2},
       adsurl = {https://ui.adsabs.harvard.edu/abs/1985Icar...61..355Z}
}

@PHDTHESIS{1984PhDT.........3T,
       author = {{Tholen}, David James},
        title = "{Asteroid Taxonomy from Cluster Analysis of Photometry.}",
       school = {University of Arizona, Tucson},
         year = 1984,
        month = sep,
       adsurl = {https://ui.adsabs.harvard.edu/abs/1984PhDT.........3T}
}

@ARTICLE{2002Icar..158..146B,
       author = {{Bus}, Schelte J. and {Binzel}, Richard P.},
        title = "{Phase II of the Small Main-Belt Asteroid Spectroscopic Survey. A Feature-Based Taxonomy}",
      journal = {Icarus},
         year = 2002,
        month = jul,
       volume = {158},
       number = {1},
        pages = {146-177},
          doi = {10.1006/icar.2002.6856},
       adsurl = {https://ui.adsabs.harvard.edu/abs/2002Icar..158..146B}
}

@ARTICLE{2009Icar..202..160D,
       author = {{DeMeo}, Francesca E. and {Binzel}, Richard P. and {Slivan}, Stephen M. and
         {Bus}, Schelte J.},
        title = "{An extension of the Bus asteroid taxonomy into the near-infrared}",
      journal = {Icarus},
         year = 2009,
        month = jul,
       volume = {202},
       number = {1},
        pages = {160-180},
          doi = {10.1016/j.icarus.2009.02.005},
       adsurl = {https://ui.adsabs.harvard.edu/abs/2009Icar..202..160D}
}

@ARTICLE{2002Icar..158..106B,
       author = {{Bus}, Schelte J. and {Binzel}, Richard P.},
        title = "{Phase II of the Small Main-Belt Asteroid Spectroscopic Survey. The Observations}",
      journal = {Icarus},
         year = 2002,
        month = jul,
       volume = {158},
       number = {1},
        pages = {106-145},
          doi = {10.1006/icar.2002.6857},
       adsurl = {https://ui.adsabs.harvard.edu/abs/2002Icar..158..106B}
}

@ARTICLE{2013Icar..226..723D,
       author = {{DeMeo}, F.~E. and {Carry}, B.},
        title = "{The taxonomic distribution of asteroids from multi-filter all-sky photometric surveys}",
      journal = {Icarus},
         year = 2013,
        month = sep,
       volume = {226},
       number = {1},
        pages = {723-741},
          doi = {10.1016/j.icarus.2013.06.027},
archivePrefix = {arXiv},
       eprint = {1307.2424},
 primaryClass = {astro-ph.EP},
       adsurl = {https://ui.adsabs.harvard.edu/abs/2013Icar..226..723D}
}

@ARTICLE{2014ApJ...791..121M,
       author = {{Masiero}, Joseph R. and {Grav}, T. and {Mainzer}, A.~K. and {Nugent}, C.~R. and {Bauer}, J.~M. and {Stevenson}, R. and {Sonnett}, S.},
        title = "{Main-belt Asteroids with WISE/NEOWISE: Near-infrared Albedos}",
      journal = {\apj},
         year = 2014,
        month = aug,
       volume = {791},
       number = {2},
          eid = {121},
        pages = {121},
          doi = {10.1088/0004-637X/791/2/121},
archivePrefix = {arXiv},
       eprint = {1406.6645},
 primaryClass = {astro-ph.EP},
       adsurl = {https://ui.adsabs.harvard.edu/abs/2014ApJ...791..121M}
}

@ARTICLE{2016Sci...353.1008R,
       author = {{Russell}, C.~T. and {Raymond}, C.~A. and {Ammannito}, E. and
         {Buczkowski}, D.~L. and {De Sanctis}, M.~C. and {Hiesinger}, H. and
         {Jaumann}, R. and {Konopliv}, A.~S. and {McSween}, H.~Y. and
         {Nathues}, A. and {Park}, R.~S. and {Pieters}, C.~M. and
         {Prettyman}, T.~H. and {McCord}, T.~B. and {McFadden}, L.~A. and
         {Mottola}, S. and {Zuber}, M.~T. and {Joy}, S.~P. and {Polanskey}, C. and
         {Rayman}, M.~D. and {Castillo-Rogez}, J.~C. and {Chi}, P.~J. and
         {Combe}, J.~P. and {Ermakov}, A. and {Fu}, R.~R. and {Hoffmann}, M. and
         {Jia}, Y.~D. and {King}, S.~D. and {Lawrence}, D.~J. and {Li}, J. -Y. and
         {Marchi}, S. and {Preusker}, F. and {Roatsch}, T. and {Ruesch}, O. and
         {Schenk}, P. and {Villarreal}, M.~N. and {Yamashita}, N.},
        title = "{Dawn arrives at Ceres: Exploration of a small, volatile-rich world}",
      journal = {Science},
         year = 2016,
        month = sep,
       volume = {353},
       number = {6303},
        pages = {1008-1010},
          doi = {10.1126/science.aaf4219},
       adsurl = {https://ui.adsabs.harvard.edu/abs/2016Sci...353.1008R}
}

@ARTICLE{2016ApJ...817L..22L,
       author = {{Li}, Jian-Yang and {Reddy}, Vishnu and {Nathues}, Andreas and {Le Corre}, Lucille and {Izawa}, Matthew R.~M. and {Cloutis}, Edward A. and {Sykes}, Mark V. and {Carsenty}, Uri and {Castillo-Rogez}, Julie C. and {Hoffmann}, Martin and {Jaumann}, Ralf and {Krohn}, Katrin and {Mottola}, Stefano and {Prettyman}, Thomas H. and {Schaefer}, Michael and {Schenk}, Paul and {Schr{\"o}der}, Stefan E. and {Williams}, David A. and {Smith}, David E. and {Zuber}, Maria T. and {Konopliv}, Alexander S. and {Park}, Ryan S. and {Raymond}, Carol A. and {Russell}, Christopher T.},
        title = "{Surface Albedo and Spectral Variability of Ceres}",
      journal = {\apjl},
         year = 2016,
        month = feb,
       volume = {817},
       number = {2},
          eid = {L22},
        pages = {L22},
          doi = {10.3847/2041-8205/817/2/L22},
archivePrefix = {arXiv},
       eprint = {1601.03713},
 primaryClass = {astro-ph.EP},
       adsurl = {https://ui.adsabs.harvard.edu/abs/2016ApJ...817L..22L}
}

@ARTICLE{2022A&A...664A.107T,
       author = {{Tatsumi}, E. and {Tinaut-Ruano}, F. and {de Le{\'o}n}, J. and {Popescu}, M. and {Licandro}, J.},
        title = "{Near-ultraviolet to visible spectroscopy of the Themis and Polana-Eulalia complex families}",
      journal = {\aap},
         year = 2022,
        month = aug,
       volume = {664},
          eid = {A107},
        pages = {A107},
          doi = {10.1051/0004-6361/202243806},
archivePrefix = {arXiv},
       eprint = {2205.13917},
 primaryClass = {astro-ph.EP},
       adsurl = {https://ui.adsabs.harvard.edu/abs/2022A&A...664A.107T}
}

@INPROCEEDINGS{2015aste.book...43R,
       author = {{Reddy}, V. and {Dunn}, T.~L. and {Thomas}, C.~A. and {Moskovitz}, N.~A. and {Burbine}, T.~H.},
        title = "{Mineralogy and Surface Composition of Asteroids}",
    booktitle = {Asteroids IV},
         year = 2015,
        pages = {43-63},
          doi = {10.2458/azu\_uapress\_9780816532131-ch003},
       adsurl = {https://ui.adsabs.harvard.edu/abs/2015aste.book...43R}
}

@ARTICLE{2021A&A...655A..47M,
       author = {{Morate}, David and {Marcio Carvano}, Jorge and {Alvarez-Candal}, Alvaro and {De Pr{\'a}}, M{\'a}rio and {Licandro}, Javier and {Galarza}, Andr{\'e}s and {Mahlke}, Max and {Solano-M{\'a}rquez}, Enrique and {Cenarro}, Javier and {Crist{\'o}bal-Hornillos}, David and {Hern{\'a}ndez-Monteagudo}, Carlos and {L{\'o}pez-Sanjuan}, Carlos and {Mar{\'\i}n-Franch}, Antonio and {Moles}, Mariano and {Varela}, Jes{\'u}s and {V{\'a}zquez Rami{\'o}}, H{\'e}ctor and {Alcaniz}, Jailson and {Dupke}, Renato and {Ederoclite}, Alessandro and {Sodr{\'e}}, Laerte and {Angulo}, Raul E. and {Jim{\'e}nez-Esteban}, Francisco M. and {Siffert}, Beatriz B. and {J-PLUS Collaboration}},
        title = "{J-PLUS: A first glimpse at the spectrophotometry of asteroids. The MOOJa catalog}",
      journal = {\aap},
         year = 2021,
        month = nov,
       volume = {655},
          eid = {A47},
        pages = {A47},
          doi = {10.1051/0004-6361/202038477},
archivePrefix = {arXiv},
       eprint = {2106.00713},
 primaryClass = {astro-ph.EP},
       adsurl = {https://ui.adsabs.harvard.edu/abs/2021A&A...655A..47M}
}

@ARTICLE{2022A&A...658A.109S,
       author = {{Sergeyev}, A.~V. and {Carry}, B. and {Onken}, C.~A. and {Devillepoix}, H.~A.~R. and {Wolf}, C. and {Chang}, S. -W.},
        title = "{Multifilter photometry of Solar System objects from the SkyMapper Southern Survey}",
      journal = {\aap},
         year = 2022,
        month = feb,
       volume = {658},
          eid = {A109},
        pages = {A109},
          doi = {10.1051/0004-6361/202142074},
archivePrefix = {arXiv},
       eprint = {2110.11656},
 primaryClass = {astro-ph.EP},
       adsurl = {https://ui.adsabs.harvard.edu/abs/2022A&A...658A.109S}
}

@Article{tanga_dr3,
  author        = {{Tanga}, P. and {Pauwels}, T. and {Mignard}, F. and {Muinonen}, K. and {Cellino}, A. and {David}, P. and {Hestroffer}, D. and {Spoto}, F. and {Berthier}, J. and {Guiraud}, J. and {Roux}, W. and {Carry}, B. and {Delbo}, M. and {Dell'Oro}, A. and {Fouron}, C. and {Galluccio}, L. and {Jonckheere}, A. and {Klioner}, S.~A. and {Lefustec}, Y. and {Liberato}, L. and {Ord{\'e}novic}, C. and {Oreshina-Slezak}, I. and {Penttil{\"a}}, A. and {Pailler}, F. and {Panem}, Ch. and {Petit}, J. -M. and {Portell}, J. and {Poujoulet}, E. and {Thuillot}, W. and {Van Hemelryck}, E. and {Burlacu}, A. and {Lasne}, Y. and {Managau}, S.},
  title         = {{Gaia Data Release 3. The Solar System survey}},
  journal       = {\aap},
  year          = {2023},
  volume        = {674},
  pages         = {A12},
  month         = jun,
  adsurl        = {https://ui.adsabs.harvard.edu/abs/2023A&A...674A..12T},
  archiveprefix = {arXiv},
  doi           = {10.1051/0004-6361/202243796},
  eid           = {A12},
  eprint        = {2206.05561},
  primaryclass  = {astro-ph.EP},
}

@ARTICLE{prusti2016,
       author = {{Prusti}, T. and {de Bruijne}, J.~H.~J. and {Brown}, A.~G.~A. and {Vallenari}, A. and {Babusiaux}, C. and {Bailer-Jones}, C.~A.~L. and {Bastian}, U. and {Biermann}, M. and {Evans}, D.~W. and {Eyer}, L. and {Jansen}, F. and {Jordi}, C. and {Klioner}, S.~A. and {Lammers}, U. and {Lindegren}, L. and {Luri}, X. and {Mignard}, F. and {Milligan}, D.~J. and {Panem}, C. and {Poinsignon}, V. and {Pourbaix}, D. and {Randich}, S. and {Sarri}, G. and {Sartoretti}, P. and {Siddiqui}, H.~I. and {Soubiran}, C. and {Valette}, V. and {van Leeuwen}, F. and {Walton}, N.~A. and {Aerts}, C. and {Arenou}, F. and {Cropper}, M. and {Drimmel}, R. and {H{\o}g}, E. and {Katz}, D. and {Lattanzi}, M.~G. and {O'Mullane}, W. and {Grebel}, E.~K. and {Holland}, A.~D. and {Huc}, C. and {Passot}, X. and {Bramante}, L. and {Cacciari}, C. and {Casta{\~n}eda}, J. and {Chaoul}, L. and {Cheek}, N. and {De Angeli}, F. and {Fabricius}, C. and {Guerra}, R. and {Hern{\'a}ndez}, J. and {Jean-Antoine-Piccolo}, A. and {Masana}, E. and {Messineo}, R. and {Mowlavi}, N. and {Nienartowicz}, K. and {Ord{\'o}{\~n}ez-Blanco}, D. and {Panuzzo}, P. and {Portell}, J. and {Richards}, P.~J. and {Riello}, M. and {Seabroke}, G.~M. and {Tanga}, P. and {Th{\'e}venin}, F. and {Torra}, J. and {Els}, S.~G. and {Gracia-Abril}, G. and {Comoretto}, G. and {Garcia-Reinaldos}, M. and {Lock}, T. and {Mercier}, E. and {Altmann}, M. and {Andrae}, R. and {Astraatmadja}, T.~L. and {Bellas-Velidis}, I. and {Benson}, K. and {Berthier}, J. and {Blomme}, R. and {Busso}, G. and {Carry}, B. and {Cellino}, A. and {Clementini}, G. and {Cowell}, S. and {Creevey}, O. and {Cuypers}, J. and {Davidson}, M. and {De Ridder}, J. and {de Torres}, A. and {Delchambre}, L. and {Dell'Oro}, A. and {Ducourant}, C. and {Fr{\'e}mat}, Y. and {Garc{\'\i}a-Torres}, M. and {Gosset}, E. and {Halbwachs}, J. -L. and {Hambly}, N.~C. and {Harrison}, D.~L. and {Hauser}, M. and {Hestroffer}, D. and {Hodgkin}, S.~T. and {Huckle}, H.~E. and {Hutton}, A. and {Jasniewicz}, G. and {Jordan}, S. and {Kontizas}, M. and {Korn}, A.~J. and {Lanzafame}, A.~C. and {Manteiga}, M. and {Moitinho}, A. and {Muinonen}, K. and {Osinde}, J. and {Pancino}, E. and {Pauwels}, T. and {Petit}, J. -M. and {Recio-Blanco}, A. and {Robin}, A.~C. and {Sarro}, L.~M. and {Siopis}, C. and {Smith}, M. and {Smith}, K.~W. and {Sozzetti}, A. and {Thuillot}, W. and {van Reeven}, W. and {Viala}, Y. and {Abbas}, U. and {Abreu Aramburu}, A. and {Accart}, S. and {Aguado}, J.~J. and {Allan}, P.~M. and {Allasia}, W. and {Altavilla}, G. and {{\'A}lvarez}, M.~A. and {Alves}, J. and {Anderson}, R.~I. and {Andrei}, A.~H. and {Anglada Varela}, E. and {Antiche}, E. and {Antoja}, T. and {Ant{\'o}n}, S. and {Arcay}, B. and {Atzei}, A. and {Ayache}, L. and {Bach}, N. and {Baker}, S.~G. and {Balaguer-N{\'u}{\~n}ez}, L. and {Barache}, C. and {Barata}, C. and {Barbier}, A. and {Barblan}, F. and {Baroni}, M. and {Barrado y Navascu{\'e}s}, D. and {Barros}, M. and {Barstow}, M.~A. and {Becciani}, U. and {Bellazzini}, M. and {Bellei}, G. and {Bello Garc{\'\i}a}, A. and {Belokurov}, V. and {Bendjoya}, P. and {Berihuete}, A. and {Bianchi}, L. and {Bienaym{\'e}}, O. and {Billebaud}, F. and {Blagorodnova}, N. and {Blanco-Cuaresma}, S. and {Boch}, T. and {Bombrun}, A. and {Borrachero}, R. and {Bouquillon}, S. and {Bourda}, G. and {Bouy}, H. and {Bragaglia}, A. and {Breddels}, M.~A. and {Brouillet}, N. and {Br{\"u}semeister}, T. and {Bucciarelli}, B. and {Budnik}, F. and {Burgess}, P. and {Burgon}, R. and {Burlacu}, A. and {Busonero}, D. and {Buzzi}, R. and {Caffau}, E. and {Cambras}, J. and {Campbell}, H. and {Cancelliere}, R. and {Cantat-Gaudin}, T. and {Carlucci}, T. and {Carrasco}, J.~M. and {Castellani}, M. and {Charlot}, P. and {Charnas}, J. and {Charvet}, P. and {Chassat}, F. and {Chiavassa}, A. and {Clotet}, M. and {Cocozza}, G. and {Collins}, R.~S. and {Collins}, P. and {Costigan}, G.},
        title = "{The Gaia mission}",
      journal = {\aap},
         year = 2016,
        month = nov,
       volume = {595},
          eid = {A1},
        pages = {A1},
          doi = {10.1051/0004-6361/201629272},
archivePrefix = {arXiv},
       eprint = {1609.04153},
 primaryClass = {astro-ph.IM},
       adsurl = {https://ui.adsabs.harvard.edu/abs/2016A&A...595A...1G}
}

@ARTICLE{2023A&A...674A..35G,
       author = {{Galluccio}, L. and {Delbo}, M. and {De Angeli}, F. and {Pauwels}, T. and {Tanga}, P. and {Mignard}, F. and {Cellino}, A. and {Brown}, A.~G.~A. and {Muinonen}, K. and {Penttil{\"a}}, A. and {Jordan}, S. and {Vallenari}, A. and {Prusti}, T. and {de Bruijne}, J.~H.~J. and {Arenou}, F. and {Babusiaux}, C. and {Biermann}, M. and {Creevey}, O.~L. and {Ducourant}, C. and {Evans}, D.~W. and {Eyer}, L. and {Guerra}, R. and {Hutton}, A. and {Jordi}, C. and {Klioner}, S.~A. and {Lammers}, U.~L. and {Lindegren}, L. and {Luri}, X. and {Panem}, C. and {Pourbaix}, D. and {Randich}, S. and {Sartoretti}, P. and {Soubiran}, C. and {Walton}, N.~A. and {Bailer-Jones}, C.~A.~L. and {Bastian}, U. and {Drimmel}, R. and {Jansen}, F. and {Katz}, D. and {Lattanzi}, M.~G. and {van Leeuwen}, F. and {Bakker}, J. and {Cacciari}, C. and {Casta{\~n}eda}, J. and {Fabricius}, C. and {Fouesneau}, M. and {Fr{\'e}mat}, Y. and {Guerrier}, A. and {Heiter}, U. and {Masana}, E. and {Messineo}, R. and {Mowlavi}, N. and {Nicolas}, C. and {Nienartowicz}, K. and {Pailler}, F. and {Panuzzo}, P. and {Riclet}, F. and {Roux}, W. and {Seabroke}, G.~M. and {Sordo}, R. and {Th{\'e}venin}, F. and {Gracia-Abril}, G. and {Portell}, J. and {Teyssier}, D. and {Altmann}, M. and {Andrae}, R. and {Audard}, M. and {Bellas-Velidis}, I. and {Benson}, K. and {Berthier}, J. and {Blomme}, R. and {Burgess}, P.~W. and {Busonero}, D. and {Busso}, G. and {C{\'a}novas}, H. and {Carry}, B. and {Cheek}, N. and {Clementini}, G. and {Damerdji}, Y. and {Davidson}, M. and {de Teodoro}, P. and {Nu{\~n}ez Campos}, M. and {Delchambre}, L. and {Dell'Oro}, A. and {Esquej}, P. and {Fern{\'a}ndez-Hern{\'a}ndez}, J. and {Fraile}, E. and {Garabato}, D. and {Garc{\'\i}a-Lario}, P. and {Gosset}, E. and {Haigron}, R. and {Halbwachs}, J. -L. and {Hambly}, N.~C. and {Harrison}, D.~L. and {Hern{\'a}ndez}, J. and {Hestroffer}, D. and {Hodgkin}, S.~T. and {Holl}, B. and {Jan{\ss}en}, K. and {Jevardat de Fombelle}, G. and {Krone-Martins}, A. and {Lanzafame}, A.~C. and {L{\"o}ffler}, W. and {Marchal}, O. and {Marrese}, P.~M. and {Moitinho}, A. and {Osborne}, P. and {Pancino}, E. and {Recio-Blanco}, A. and {Reyl{\'e}}, C. and {Riello}, M. and {Rimoldini}, L. and {Roegiers}, T. and {Rybizki}, J. and {Sarro}, L.~M. and {Siopis}, C. and {Smith}, M. and {Sozzetti}, A. and {Utrilla}, E. and {van Leeuwen}, M. and {Abbas}, U. and {{\'A}brah{\'a}m}, P. and {Abreu Aramburu}, A. and {Aerts}, C. and {Aguado}, J.~J. and {Ajaj}, M. and {Aldea-Montero}, F. and {Altavilla}, G. and {{\'A}lvarez}, M.~A. and {Alves}, J. and {Anderson}, R.~I. and {Anglada Varela}, E. and {Antoja}, T. and {Baines}, D. and {Baker}, S.~G. and {Balaguer-N{\'u}{\~n}ez}, L. and {Balbinot}, E. and {Balog}, Z. and {Barache}, C. and {Barbato}, D. and {Barros}, M. and {Barstow}, M.~A. and {Bartolom{\'e}}, S. and {Bassilana}, J. -L. and {Bauchet}, N. and {Becciani}, U. and {Bellazzini}, M. and {Berihuete}, A. and {Bernet}, M. and {Bertone}, S. and {Bianchi}, L. and {Binnenfeld}, A. and {Blanco-Cuaresma}, S. and {Boch}, T. and {Bombrun}, A. and {Bossini}, D. and {Bouquillon}, S. and {Bragaglia}, A. and {Bramante}, L. and {Breedt}, E. and {Bressan}, A. and {Brouillet}, N. and {Brugaletta}, E. and {Bucciarelli}, B. and {Burlacu}, A. and {Butkevich}, A.~G. and {Buzzi}, R. and {Caffau}, E. and {Cancelliere}, R. and {Cantat-Gaudin}, T. and {Carballo}, R. and {Carlucci}, T. and {Carnerero}, M.~I. and {Carrasco}, J.~M. and {Casamiquela}, L. and {Castellani}, M. and {Castro-Ginard}, A. and {Chaoul}, L. and {Charlot}, P. and {Chemin}, L. and {Chiaramida}, V. and {Chiavassa}, A. and {Chornay}, N. and {Comoretto}, G. and {Contursi}, G. and {Cooper}, W.~J. and {Cornez}, T. and {Cowell}, S. and {Crifo}, F. and {Cropper}, M. and {Crosta}, M. and {Crowley}, C. and {Dafonte}, C. and {Dapergolas}, A. and {David}, P. and {de Laverny}, P. and {De Luise}, F. and {De March}, R. and {De Ridder}, J. and {de Souza}, R. and {de Torres}, A. and {del Peloso}, E.~F. and {del Pozo}, E. and {Delgado}, A. and {Delisle}, J. -B. and {Demouchy}, C. and {Dharmawardena}, T.~E. and {Diakite}, S. and {Diener}, C. and {Distefano}, E. and {Dolding}, C. and {Enke}, H. and {Fabre}, C. and {Fabrizio}, M. and {Faigler}, S. and {Fedorets}, G. and {Fernique}, P. and {Figueras}, F. and {Fournier}, Y. and {Fouron}, C. and {Fragkoudi}, F. and {Gai}, M. and {Garcia-Gutierrez}, A. and {Garcia-Reinaldos}, M. and {Garc{\'\i}a-Torres}, M. and {Garofalo}, A. and {Gavel}, A. and {Gavras}, P. and {Gerlach}, E. and {Geyer}, R. and {Giacobbe}, P. and {Gilmore}, G. and {Girona}, S. and {Giuffrida}, G. and {Gomel}, R. and {Gomez}, A. and {Gonz{\'a}lez-N{\'u}{\~n}ez}, J. and {Gonz{\'a}lez-Santamar{\'\i}a}, I. and {Gonz{\'a}lez-Vidal}, J.~J. and {Granvik}, M. and {Guillout}, P. and {Guiraud}, J. and {Guti{\'e}rrez-S{\'a}nchez}, R. and {Guy}, L.~P. and {Hatzidimitriou}, D. and {Hauser}, M. and {Haywood}, M. and {Helmer}, A. and {Helmi}, A. and {Sarmiento}, M.~H. and {Hidalgo}, S.~L. and {H{\l}adczuk}, N. and {Hobbs}, D. and {Holland}, G. and {Huckle}, H.~E. and {Jardine}, K. and {Jasniewicz}, G. and {Jean-Antoine Piccolo}, A. and {Jim{\'e}nez-Arranz}, {\'O}. and {Juaristi Campillo}, J. and {Julbe}, F. and {Karbevska}, L. and {Kervella}, P. and {Khanna}, S. and {Kordopatis}, G. and {Korn}, A.~J. and {K{\'o}sp{\'a}l}, {\'A}. and {Kostrzewa-Rutkowska}, Z. and {Kruszy{\'n}ska}, K. and {Kun}, M. and {Laizeau}, P. and {Lambert}, S. and {Lanza}, A.~F. and {Lasne}, Y. and {Le Campion}, J. -F. and {Lebreton}, Y. and {Lebzelter}, T. and {Leccia}, S. and {Leclerc}, N. and {Lecoeur-Taibi}, I. and {Liao}, S. and {Licata}, E.~L. and {Lindstr{\o}m}, H.~E.~P. and {Lister}, T.~A. and {Livanou}, E. and {Lobel}, A. and {Lorca}, A. and {Loup}, C. and {Madrero Pardo}, P. and {Magdaleno Romeo}, A. and {Managau}, S. and {Mann}, R.~G. and {Manteiga}, M. and {Marchant}, J.~M. and {Marconi}, M. and {Marcos}, J. and {Marcos Santos}, M.~M.~S. and {Mar{\'\i}n Pina}, D. and {Marinoni}, S. and {Marocco}, F. and {Marshall}, D.~J. and {Martin Polo}, L. and {Mart{\'\i}n-Fleitas}, J.~M. and {Marton}, G. and {Mary}, N. and {Masip}, A. and {Massari}, D. and {Mastrobuono-Battisti}, A. and {Mazeh}, T. and {McMillan}, P.~J. and {Messina}, S. and {Michalik}, D. and {Millar}, N.~R. and {Mints}, A. and {Molina}, D. and {Molinaro}, R. and {Moln{\'a}r}, L. and {Monari}, G. and {Mongui{\'o}}, M. and {Montegriffo}, P. and {Montero}, A. and {Mor}, R. and {Mora}, A. and {Morbidelli}, R. and {Morel}, T. and {Morris}, D. and {Muraveva}, T. and {Murphy}, C.~P. and {Musella}, I. and {Nagy}, Z. and {Noval}, L. and {Oca{\~n}a}, F. and {Ogden}, A. and {Ordenovic}, C. and {Osinde}, J.~O. and {Pagani}, C. and {Pagano}, I. and {Palaversa}, L. and {Palicio}, P.~A. and {Pallas-Quintela}, L. and {Panahi}, A. and {Payne-Wardenaar}, S. and {Pe{\~n}alosa Esteller}, X. and {Petit}, J. -M. and {Pichon}, B. and {Piersimoni}, A.~M. and {Pineau}, F. -X. and {Plachy}, E. and {Plum}, G. and {Poggio}, E. and {Pr{\v{s}}a}, A. and {Pulone}, L. and {Racero}, E. and {Ragaini}, S. and {Rainer}, M. and {Raiteri}, C.~M. and {Ramos}, P. and {Ramos-Lerate}, M. and {Re Fiorentin}, P. and {Regibo}, S. and {Richards}, P.~J. and {Rios Diaz}, C. and {Ripepi}, V. and {Riva}, A. and {Rix}, H. -W. and {Rixon}, G. and {Robichon}, N. and {Robin}, A.~C. and {Robin}, C. and {Roelens}, M. and {Rogues}, H.~R.~O. and {Rohrbasser}, L. and {Romero-G{\'o}mez}, M. and {Rowell}, N. and {Royer}, F. and {Ruz Mieres}, D. and {Rybicki}, K.~A. and {Sadowski}, G. and {S{\'a}ez N{\'u}{\~n}ez}, A. and {Sagrist{\`a} Sell{\'e}s}, A. and {Sahlmann}, J. and {Salguero}, E. and {Samaras}, N. and {Sanchez Gimenez}, V. and {Sanna}, N. and {Santove{\~n}a}, R. and {Sarasso}, M. and {Schultheis}, M. and {Sciacca}, E. and {Segol}, M. and {Segovia}, J.~C. and {S{\'e}gransan}, D. and {Semeux}, D. and {Shahaf}, S. and {Siddiqui}, H.~I. and {Siebert}, A. and {Siltala}, L. and {Silvelo}, A. and {Slezak}, E. and {Slezak}, I. and {Smart}, R.~L. and {Snaith}, O.~N. and {Solano}, E. and {Solitro}, F. and {Souami}, D. and {Souchay}, J. and {Spagna}, A. and {Spina}, L. and {Spoto}, F. and {Steele}, I.~A. and {Steidelm{\"u}ller}, H. and {Stephenson}, C.~A. and {S{\"u}veges}, M. and {Surdej}, J. and {Szabados}, L. and {Szegedi-Elek}, E. and {Taris}, F. and {Taylor}, M.~B. and {Teixeira}, R. and {Tolomei}, L. and {Tonello}, N. and {Torra}, F. and {Torra}, J. and {Torralba Elipe}, G. and {Trabucchi}, M. and {Tsounis}, A.~T. and {Turon}, C. and {Ulla}, A. and {Unger}, N. and {Vaillant}, M.~V. and {van Dillen}, E. and {van Reeven}, W. and {Vanel}, O. and {Vecchiato}, A. and {Viala}, Y. and {Vicente}, D. and {Voutsinas}, S. and {Weiler}, M. and {Wevers}, T. and {Wyrzykowski}, {\L}. and {Yoldas}, A. and {Yvard}, P. and {Zhao}, H. and {Zorec}, J. and {Zucker}, S. and {Zwitter}, T.},
        title = "{Gaia Data Release 3. Reflectance spectra of Solar System small bodies}",
      journal = {\aap},
         year = 2023,
        month = jun,
       volume = {674},
          eid = {A35},
        pages = {A35},
          doi = {10.1051/0004-6361/202243791},
archivePrefix = {arXiv},
       eprint = {2206.12174},
 primaryClass = {astro-ph.EP},
       adsurl = {https://ui.adsabs.harvard.edu/abs/2023A&A...674A..35G}
}

@ARTICLE{1951ApJ...114..522J,
       author = {{Johnson}, H.~L. and {Morgan}, W.~W.},
        title = "{On the Color-Magnitude Diagram of the Pleiades.}",
      journal = {\apj},
         year = 1951,
        month = nov,
       volume = {114},
        pages = {522},
          doi = {10.1086/145496},
       adsurl = {https://ui.adsabs.harvard.edu/abs/1951ApJ...114..522J}
}

@ARTICLE{2023A&A...669L..14T,
       author = {{Tinaut-Ruano}, F. and {Tatsumi}, E. and {Tanga}, P. and {de Le{\'o}n}, J. and {Delbo}, M. and {De Angeli}, F. and {Morate}, D. and {Licandro}, J. and {Galluccio}, L.},
        title = "{Asteroids' reflectance from Gaia DR3: Artificial reddening at near-UV wavelengths}",
      journal = {\aap},
         year = 2023,
        month = jan,
       volume = {669},
          eid = {L14},
        pages = {L14},
          doi = {10.1051/0004-6361/202245134},
archivePrefix = {arXiv},
       eprint = {2301.02157},
 primaryClass = {astro-ph.EP},
       adsurl = {https://ui.adsabs.harvard.edu/abs/2023A&A...669L..14T}
}

@ARTICLE{2021A&A...652A..59S,
       author = {{Sergeyev}, Alexey V. and {Carry}, Benoit},
        title = "{A million asteroid observations in the Sloan Digital Sky Survey}",
      journal = {\aap},
         year = 2021,
        month = aug,
       volume = {652},
          eid = {A59},
        pages = {A59},
          doi = {10.1051/0004-6361/202140430},
archivePrefix = {arXiv},
       eprint = {2108.05749},
 primaryClass = {astro-ph.EP},
       adsurl = {https://ui.adsabs.harvard.edu/abs/2021A&A...652A..59S}
}

@ARTICLE{2023A&A...671A.151B,
       author = {{Berthier}, J. and {Carry}, B. and {Mahlke}, M. and {Normand}, J.},
        title = "{SsODNet: Solar system Open Database Network}",
      journal = {\aap},
         year = 2023,
        month = mar,
       volume = {671},
          eid = {A151},
        pages = {A151},
          doi = {10.1051/0004-6361/202244878},
archivePrefix = {arXiv},
       eprint = {2209.10697},
 primaryClass = {astro-ph.EP},
       adsurl = {https://ui.adsabs.harvard.edu/abs/2023A&A...671A.151B}
}

@ARTICLE{2022A&A...665A..26M,
       author = {{Mahlke}, M. and {Carry}, B. and {Mattei}, P. -A.},
        title = "{Asteroid taxonomy from cluster analysis of spectrometry and albedo}",
      journal = {\aap},
         year = 2022,
        month = sep,
       volume = {665},
          eid = {A26},
        pages = {A26},
          doi = {10.1051/0004-6361/202243587},
archivePrefix = {arXiv},
       eprint = {2203.11229},
 primaryClass = {astro-ph.EP},
       adsurl = {https://ui.adsabs.harvard.edu/abs/2022A&A...665A..26M}
}

@ARTICLE{1999Icar..140...17T,
       author = {{Thomas}, P.~C. and {Veverka}, J. and {Bell}, J.~F. and {Clark}, B.~E. and {Carcich}, B. and {Joseph}, J. and {Robinson}, M. and {McFadden}, L.~A. and {Malin}, M.~C. and {Chapman}, C.~R. and {Merline}, W. and {Murchie}, S.},
        title = "{Mathilde: Size, Shape, and Geology}",
      journal = {\icarus},
         year = 1999,
        month = jul,
       volume = {140},
       number = {1},
        pages = {17-27},
          doi = {10.1006/icar.1999.6121},
       adsurl = {https://ui.adsabs.harvard.edu/abs/1999Icar..140...17T}
}

@ARTICLE{2002AJ....123.1056T,
       author = {{Tedesco}, Edward F. and {Noah}, Paul V. and {Noah}, Meg and {Price}, Stephan D.},
        title = "{The Supplemental IRAS Minor Planet Survey}",
      journal = {\aj},
         year = 2002,
        month = feb,
       volume = {123},
       number = {2},
        pages = {1056-1085},
          doi = {10.1086/338320},
       adsurl = {https://ui.adsabs.harvard.edu/abs/2002AJ....123.1056T}
}

@PHDTHESIS{2004PhDT.......371D,
       author = {{Delbo}, Marco},
        title = "{The nature of near-earth asteroids from the study of their thermal infrared emission}",
       school = {Free University of Berlin, Germany},
         year = 2004,
        month = jan,
       adsurl = {https://ui.adsabs.harvard.edu/abs/2004PhDT.......371D}
}

@ARTICLE{2006Icar..185...39M,
       author = {{Marchis}, F. and {Kaasalainen}, M. and {Hom}, E.~F.~Y. and {Berthier}, J. and {Enriquez}, J. and {Hestroffer}, D. and {Le Mignant}, D. and {de Pater}, I.},
        title = "{Shape, size and multiplicity of main-belt asteroids. I. Keck Adaptive Optics survey}",
      journal = {\icarus},
         year = 2006,
        month = nov,
       volume = {185},
       number = {1},
        pages = {39-63},
          doi = {10.1016/j.icarus.2006.06.001},
       adsurl = {https://ui.adsabs.harvard.edu/abs/2006Icar..185...39M}
}

@ARTICLE{2007Icar..186..152M,
       author = {{Magri}, Christopher and {Ostro}, Steven J. and {Scheeres}, Daniel J. and {Nolan}, Michael C. and {Giorgini}, Jon D. and {Benner}, Lance A.~M. and {Margot}, Jean-Luc},
        title = "{Radar observations and a physical model of Asteroid 1580 Betulia}",
      journal = {\icarus},
         year = 2007,
        month = jan,
       volume = {186},
       number = {1},
        pages = {152-177},
          doi = {10.1016/j.icarus.2006.08.004},
       adsurl = {https://ui.adsabs.harvard.edu/abs/2007Icar..186..152M}
}

@ARTICLE{2007Icar..186..126M,
       author = {{Magri}, Christopher and {Nolan}, Michael C. and {Ostro}, Steven J. and {Giorgini}, Jon D.},
        title = "{A radar survey of main-belt asteroids: Arecibo observations of 55 objects during 1999 2003}",
      journal = {\icarus},
         year = 2007,
        month = jan,
       volume = {186},
       number = {1},
        pages = {126-151},
          doi = {10.1016/j.icarus.2006.08.018},
       adsurl = {https://ui.adsabs.harvard.edu/abs/2007Icar..186..126M}
}

@ARTICLE{2008Icar..193...20S,
       author = {{Shepard}, Michael K. and {Clark}, Beth Ellen and {Nolan}, Michael C. and {Benner}, Lance A.~M. and {Ostro}, Steven J. and {Giorgini}, Jon D. and {Vilas}, Faith and {Jarvis}, Kandy and {Lederer}, Susan and {Lim}, Lucy F. and {McConnochie}, Tim and {Bell}, James and {Margot}, Jean-Luc and {Rivkin}, Andrew and {Magri}, Christopher and {Scheeres}, Daniel and {Pravec}, Petr},
        title = "{Multi-wavelength observations of Asteroid 2100 Ra-Shalom}",
      journal = {\icarus},
         year = 2008,
        month = jan,
       volume = {193},
       number = {1},
        pages = {20-38},
          doi = {10.1016/j.icarus.2007.09.006},
       adsurl = {https://ui.adsabs.harvard.edu/abs/2008Icar..193...20S}
}

@ARTICLE{2009P&SS...57..259D,
       author = {{Delbo'}, Marco and {Tanga}, Paolo},
        title = "{Thermal inertia of main belt asteroids smaller than 100 km from IRAS data}",
      journal = {\planss},
         year = 2009,
        month = feb,
       volume = {57},
       number = {2},
        pages = {259-265},
          doi = {10.1016/j.pss.2008.06.015},
archivePrefix = {arXiv},
       eprint = {0808.0869},
 primaryClass = {astro-ph},
       adsurl = {https://ui.adsabs.harvard.edu/abs/2009P&SS...57..259D}
}

@ARTICLE{2010AJ....140..770T,
       author = {{Trilling}, D.~E. and {Mueller}, M. and {Hora}, J.~L. and {Harris}, A.~W. and {Bhattacharya}, B. and {Bottke}, W.~F. and {Chesley}, S. and {Delbo}, M. and {Emery}, J.~P. and {Fazio}, G. and {Mainzer}, A. and {Penprase}, B. and {Smith}, H.~A. and {Spahr}, T.~B. and {Stansberry}, J.~A. and {Thomas}, C.~A.},
        title = "{ExploreNEOs. I. Description and First Results from the Warm Spitzer Near-Earth Object Survey}",
      journal = {\aj},
         year = 2010,
        month = sep,
       volume = {140},
       number = {3},
        pages = {770-784},
          doi = {10.1088/0004-6256/140/3/770},
       adsurl = {https://ui.adsabs.harvard.edu/abs/2010AJ....140..770T}
}

@ARTICLE{2010AJ....140..933R,
       author = {{Ryan}, Erin Lee and {Woodward}, Charles E.},
        title = "{Rectified Asteroid Albedos and Diameters from IRAS and MSX Photometry Catalogs}",
      journal = {\aj},
         year = 2010,
        month = oct,
       volume = {140},
       number = {4},
        pages = {933-943},
          doi = {10.1088/0004-6256/140/4/933},
archivePrefix = {arXiv},
       eprint = {1006.4362},
 primaryClass = {astro-ph.EP},
       adsurl = {https://ui.adsabs.harvard.edu/abs/2010AJ....140..933R}
}

@ARTICLE{2011ApJ...741...68M,
       author = {{Masiero}, Joseph R. and {Mainzer}, A.~K. and {Grav}, T. and {Bauer}, J.~M. and {Cutri}, R.~M. and {Dailey}, J. and {Eisenhardt}, P.~R.~M. and {McMillan}, R.~S. and {Spahr}, T.~B. and {Skrutskie}, M.~F. and {Tholen}, D. and {Walker}, R.~G. and {Wright}, E.~L. and {DeBaun}, E. and {Elsbury}, D. and {Gautier}, T., IV and {Gomillion}, S. and {Wilkins}, A.},
        title = "{Main Belt Asteroids with WISE/NEOWISE. I. Preliminary Albedos and Diameters}",
      journal = {\apj},
         year = 2011,
        month = nov,
       volume = {741},
       number = {2},
          eid = {68},
        pages = {68},
          doi = {10.1088/0004-637X/741/2/68},
archivePrefix = {arXiv},
       eprint = {1109.4096},
 primaryClass = {astro-ph.EP},
       adsurl = {https://ui.adsabs.harvard.edu/abs/2011ApJ...741...68M}
}

@ARTICLE{2011ApJ...742...40G,
       author = {{Grav}, T. and {Mainzer}, A.~K. and {Bauer}, J. and {Masiero}, J. and {Spahr}, T. and {McMillan}, R.~S. and {Walker}, R. and {Cutri}, R. and {Wright}, E. and {Eisenhardt}, P.~R.~M. and {Blauvelt}, E. and {DeBaun}, E. and {Elsbury}, D. and {Gautier}, T., IV and {Gomillion}, S. and {Hand}, E. and {Wilkins}, A.},
        title = "{WISE/NEOWISE Observations of the Jovian Trojans: Preliminary Results}",
      journal = {\apj},
         year = 2011,
        month = nov,
       volume = {742},
       number = {1},
          eid = {40},
        pages = {40},
          doi = {10.1088/0004-637X/742/1/40},
archivePrefix = {arXiv},
       eprint = {1110.0280},
 primaryClass = {astro-ph.EP},
       adsurl = {https://ui.adsabs.harvard.edu/abs/2011ApJ...742...40G}
}

@ARTICLE{2011ApJ...743..156M,
       author = {{Mainzer}, A. and {Grav}, T. and {Bauer}, J. and {Masiero}, J. and {McMillan}, R.~S. and {Cutri}, R.~M. and {Walker}, R. and {Wright}, E. and {Eisenhardt}, P. and {Tholen}, D.~J. and {Spahr}, T. and {Jedicke}, R. and {Denneau}, L. and {DeBaun}, E. and {Elsbury}, D. and {Gautier}, T. and {Gomillion}, S. and {Hand}, E. and {Mo}, W. and {Watkins}, J. and {Wilkins}, A. and {Bryngelson}, G.~L. and {Del Pino Molina}, A. and {Desai}, S. and {G{\'o}mez Camus}, M. and {Hidalgo}, S.~L. and {Konstantopoulos}, I. and {Larsen}, J.~A. and {Maleszewski}, C. and {Malkan}, M.~A. and {Mauduit}, J. -C. and {Mullan}, B.~L. and {Olszewski}, E.~W. and {Pforr}, J. and {Saro}, A. and {Scotti}, J.~V. and {Wasserman}, L.~H.},
        title = "{NEOWISE Observations of Near-Earth Objects: Preliminary Results}",
      journal = {\apj},
         year = 2011,
        month = dec,
       volume = {743},
       number = {2},
          eid = {156},
        pages = {156},
          doi = {10.1088/0004-637X/743/2/156},
archivePrefix = {arXiv},
       eprint = {1109.6400},
 primaryClass = {astro-ph.EP},
       adsurl = {https://ui.adsabs.harvard.edu/abs/2011ApJ...743..156M}
}

@ARTICLE{2011Icar..214..652D,
       author = {{{\v{D}}urech}, Josef and {Kaasalainen}, Mikko and {Herald}, David and {Dunham}, David and {Timerson}, Brad and {Hanu{\v{s}}}, Josef and {Frappa}, Eric and {Talbot}, John and {Hayamizu}, Tsutomu and {Warner}, Brian D. and {Pilcher}, Frederick and {Gal{\'a}d}, Adri{\'a}n},
        title = "{Combining asteroid models derived by lightcurve inversion with asteroidal occultation silhouettes}",
      journal = {\icarus},
         year = 2011,
        month = aug,
       volume = {214},
       number = {2},
        pages = {652-670},
          doi = {10.1016/j.icarus.2011.03.016},
archivePrefix = {arXiv},
       eprint = {1104.4227},
 primaryClass = {astro-ph.EP},
       adsurl = {https://ui.adsabs.harvard.edu/abs/2011Icar..214..652D}
}

@ARTICLE{2011PASJ...63.1117U,
       author = {{Usui}, Fumihiko and {Kuroda}, Daisuke and {M{\"u}ller}, Thomas G. and {Hasegawa}, Sunao and {Ishiguro}, Masateru and {Ootsubo}, Takafumi and {Ishihara}, Daisuke and {Kataza}, Hirokazu and {Takita}, Satoshi and {Oyabu}, Shinki and {Ueno}, Munetaka and {Matsuhara}, Hideo and {Onaka}, Takashi},
        title = "{Asteroid Catalog Using Akari: AKARI/IRC Mid-Infrared Asteroid Survey}",
      journal = {\pasj},
         year = 2011,
        month = oct,
       volume = {63},
        pages = {1117-1138},
          doi = {10.1093/pasj/63.5.1117},
       adsurl = {https://ui.adsabs.harvard.edu/abs/2011PASJ...63.1117U}
}

@ARTICLE{2011AJ....141..186R,
       author = {{Ryan}, Erin Lee and {Woodward}, Charles E.},
        title = "{Albedos of Small Hilda Group Asteroids as Revealed by Spitzer}",
      journal = {\aj},
         year = 2011,
        month = jun,
       volume = {141},
       number = {6},
          eid = {186},
        pages = {186},
          doi = {10.1088/0004-6256/141/6/186},
archivePrefix = {arXiv},
       eprint = {1103.0724},
 primaryClass = {astro-ph.EP},
       adsurl = {https://ui.adsabs.harvard.edu/abs/2011AJ....141..186R}
}

@ARTICLE{2012ApJ...744..197G,
       author = {{Grav}, T. and {Mainzer}, A.~K. and {Bauer}, J. and {Masiero}, J. and {Spahr}, T. and {McMillan}, R.~S. and {Walker}, R. and {Cutri}, R. and {Wright}, E. and {Eisenhardt}, P.~R. and {Blauvelt}, E. and {DeBaun}, E. and {Elsbury}, D. and {Gautier}, T. and {Gomillion}, S. and {Hand}, E. and {Wilkins}, A.},
        title = "{WISE/NEOWISE Observations of the Hilda Population: Preliminary Results}",
      journal = {\apj},
         year = 2012,
        month = jan,
       volume = {744},
       number = {2},
          eid = {197},
        pages = {197},
          doi = {10.1088/0004-637X/744/2/197},
archivePrefix = {arXiv},
       eprint = {1110.0283},
 primaryClass = {astro-ph.EP},
       adsurl = {https://ui.adsabs.harvard.edu/abs/2012ApJ...744..197G}
}

@ARTICLE{2012ApJ...759...49G,
       author = {{Grav}, T. and {Mainzer}, A.~K. and {Bauer}, J.~M. and {Masiero}, J.~R. and {Nugent}, C.~R.},
        title = "{WISE/NEOWISE Observations of the Jovian Trojan Population: Taxonomy}",
      journal = {\apj},
         year = 2012,
        month = nov,
       volume = {759},
       number = {1},
          eid = {49},
        pages = {49},
          doi = {10.1088/0004-637X/759/1/49},
archivePrefix = {arXiv},
       eprint = {1209.1549},
 primaryClass = {astro-ph.EP},
       adsurl = {https://ui.adsabs.harvard.edu/abs/2012ApJ...759...49G}
}

@ARTICLE{2012ApJ...759L...8M,
       author = {{Masiero}, Joseph R. and {Mainzer}, A.~K. and {Grav}, T. and {Bauer}, J.~M. and {Cutri}, R.~M. and {Nugent}, C. and {Cabrera}, M.~S.},
        title = "{Preliminary Analysis of WISE/NEOWISE 3-Band Cryogenic and Post-cryogenic Observations of Main Belt Asteroids}",
      journal = {\apjl},
         year = 2012,
        month = nov,
       volume = {759},
       number = {1},
          eid = {L8},
        pages = {L8},
          doi = {10.1088/2041-8205/759/1/L8},
archivePrefix = {arXiv},
       eprint = {1209.5794},
 primaryClass = {astro-ph.EP},
       adsurl = {https://ui.adsabs.harvard.edu/abs/2012ApJ...759L...8M}
}

@ARTICLE{2012Icar..221..365P,
       author = {{Pravec}, Petr and {Harris}, Alan W. and {Ku{\v{s}}nir{\'a}k}, Peter and {Gal{\'a}d}, Adri{\'a}n and {Hornoch}, Kamil},
        title = "{Absolute magnitudes of asteroids and a revision of asteroid albedo estimates from WISE thermal observations}",
      journal = {\icarus},
         year = 2012,
        month = sep,
       volume = {221},
       number = {1},
        pages = {365-387},
          doi = {10.1016/j.icarus.2012.07.026},
       adsurl = {https://ui.adsabs.harvard.edu/abs/2012Icar..221..365P}
}

@ARTICLE{2012Icar..221.1130M,
       author = {{Marchis}, F. and {Enriquez}, J.~E. and {Emery}, J.~P. and {Mueller}, M. and {Baek}, M. and {Pollock}, J. and {Assafin}, M. and {Vieira Martins}, R. and {Berthier}, J. and {Vachier}, F. and {Cruikshank}, D.~P. and {Lim}, L.~F. and {Reichart}, D.~E. and {Ivarsen}, K.~M. and {Haislip}, J.~B. and {LaCluyze}, A.~P.},
        title = "{Multiple asteroid systems: Dimensions and thermal properties from Spitzer Space Telescope and ground-based observations}",
      journal = {\icarus},
         year = 2012,
        month = nov,
       volume = {221},
       number = {2},
        pages = {1130-1161},
          doi = {10.1016/j.icarus.2012.09.013},
archivePrefix = {arXiv},
       eprint = {1604.05384},
 primaryClass = {astro-ph.EP},
       adsurl = {https://ui.adsabs.harvard.edu/abs/2012Icar..221.1130M}
}

@ARTICLE{2012MNRAS.423.2587H,
       author = {{Horner}, J. and {M{\"u}ller}, T.~G. and {Lykawka}, P.~S.},
        title = "{(1173) Anchises - thermophysical and dynamical studies of a dynamically unstable Jovian Trojan}",
      journal = {\mnras},
         year = 2012,
        month = jul,
       volume = {423},
       number = {3},
        pages = {2587-2596},
          doi = {10.1111/j.1365-2966.2012.21067.x},
archivePrefix = {arXiv},
       eprint = {1204.1388},
 primaryClass = {astro-ph.EP},
       adsurl = {https://ui.adsabs.harvard.edu/abs/2012MNRAS.423.2587H}
}

@ARTICLE{2012ApJ...760L..12M,
       author = {{Mainzer}, A. and {Grav}, T. and {Masiero}, J. and {Bauer}, J. and {Cutri}, R.~M. and {McMillan}, R.~S. and {Nugent}, C.~R. and {Tholen}, D. and {Walker}, R. and {Wright}, E.~L.},
        title = "{Physical Parameters of Asteroids Estimated from the WISE 3-Band Data and NEOWISE Post-Cryogenic Survey}",
      journal = {\apjl},
         year = 2012,
        month = nov,
       volume = {760},
       number = {1},
          eid = {L12},
        pages = {L12},
          doi = {10.1088/2041-8205/760/1/L12},
archivePrefix = {arXiv},
       eprint = {1210.0502},
 primaryClass = {astro-ph.EP},
       adsurl = {https://ui.adsabs.harvard.edu/abs/2012ApJ...760L..12M}
}

@ARTICLE{2013A&A...554A..71A,
       author = {{Al{\'\i}-Lagoa}, V. and {de Le{\'o}n}, J. and {Licandro}, J. and {Delb{\'o}}, M. and {Campins}, H. and {Pinilla-Alonso}, N. and {Kelley}, M.~S.},
        title = "{Physical properties of B-type asteroids from WISE data}",
      journal = {\aap},
         year = 2013,
        month = jun,
       volume = {554},
          eid = {A71},
        pages = {A71},
          doi = {10.1051/0004-6361/201220680},
archivePrefix = {arXiv},
       eprint = {1303.5487},
 primaryClass = {astro-ph.EP},
       adsurl = {https://ui.adsabs.harvard.edu/abs/2013A&A...554A..71A}
}

@ARTICLE{2013A&A...555A..15F,
       author = {{Fornasier}, S. and {Lellouch}, E. and {M{\"u}ller}, T. and {Santos-Sanz}, P. and {Panuzzo}, P. and {Kiss}, C. and {Lim}, T. and {Mommert}, M. and {Bockel{\'e}e-Morvan}, D. and {Vilenius}, E. and {Stansberry}, J. and {Tozzi}, G.~P. and {Mottola}, S. and {Delsanti}, A. and {Crovisier}, J. and {Duffard}, R. and {Henry}, F. and {Lacerda}, P. and {Barucci}, A. and {Gicquel}, A.},
        title = "{TNOs are Cool: A survey of the trans-Neptunian region. VIII. Combined Herschel PACS and SPIRE observations of nine bright targets at 70-500 {\ensuremath{\mu}}m}",
      journal = {\aap},
         year = 2013,
        month = jul,
       volume = {555},
          eid = {A15},
        pages = {A15},
          doi = {10.1051/0004-6361/201321329},
archivePrefix = {arXiv},
       eprint = {1305.0449},
 primaryClass = {astro-ph.EP},
       adsurl = {https://ui.adsabs.harvard.edu/abs/2013A&A...555A..15F}
}

@ARTICLE{2013ApJ...773...22B,
       author = {{Bauer}, James M. and {Grav}, Tommy and {Blauvelt}, Erin and {Mainzer}, A.~K. and {Masiero}, Joseph R. and {Stevenson}, Rachel and {Kramer}, Emily and {Fern{\'a}ndez}, Yan R. and {Lisse}, C.~M. and {Cutri}, Roc M. and {Weissman}, Paul R. and {Dailey}, John W. and {Masci}, Frank J. and {Walker}, Russel and {Waszczak}, Adam and {Nugent}, Carrie R. and {Meech}, Karen J. and {Lucas}, Andrew and {Pearman}, George and {Wilkins}, Ashlee and {Watkins}, Jessica and {Kulkarni}, Shrinivas and {Wright}, Edward L. and {WISE Team} and {PTF Team}},
        title = "{Centaurs and Scattered Disk Objects in the Thermal Infrared: Analysis of WISE/NEOWISE Observations}",
      journal = {\apj},
         year = 2013,
        month = aug,
       volume = {773},
       number = {1},
          eid = {22},
        pages = {22},
          doi = {10.1088/0004-637X/773/1/22},
archivePrefix = {arXiv},
       eprint = {1306.1862},
 primaryClass = {astro-ph.EP},
       adsurl = {https://ui.adsabs.harvard.edu/abs/2013ApJ...773...22B}
}

@ARTICLE{2013Icar..226.1045H,
       author = {{Hanu{\v{s}}}, J. and {Marchis}, F. and {{\v{D}}urech}, J.},
        title = "{Sizes of main-belt asteroids by combining shape models and Keck adaptive optics observations}",
      journal = {\icarus},
         year = 2013,
        month = sep,
       volume = {226},
       number = {1},
        pages = {1045-1057},
          doi = {10.1016/j.icarus.2013.07.023},
archivePrefix = {arXiv},
       eprint = {1308.0446},
 primaryClass = {astro-ph.EP},
       adsurl = {https://ui.adsabs.harvard.edu/abs/2013Icar..226.1045H}
}

@ARTICLE{2013PASJ...65...34H,
       author = {{Hasegawa}, Sunao and {M{\"u}ller}, Thomas G. and {Kuroda}, Daisuke and {Takita}, Satoshi and {Usui}, Fumihiko},
        title = "{The Asteroid Catalog Using AKARI IRC Slow-Scan Observations}",
      journal = {\pasj},
         year = 2013,
        month = apr,
       volume = {65},
          eid = {34},
        pages = {34},
          doi = {10.1093/pasj/65.2.34},
archivePrefix = {arXiv},
       eprint = {1210.7557},
 primaryClass = {astro-ph.EP},
       adsurl = {https://ui.adsabs.harvard.edu/abs/2013PASJ...65...34H}
}

@ARTICLE{2014A&A...564A..92D,
       author = {{Duffard}, R. and {Pinilla-Alonso}, N. and {Santos-Sanz}, P. and {Vilenius}, E. and {Ortiz}, J.~L. and {Mueller}, T. and {Fornasier}, S. and {Lellouch}, E. and {Mommert}, M. and {Pal}, A. and {Kiss}, C. and {Mueller}, M. and {Stansberry}, J. and {Delsanti}, A. and {Peixinho}, N. and {Trilling}, D.},
        title = "{``TNOs are Cool'': A survey of the trans-Neptunian region. XI. A Herschel-PACS view of 16 Centaurs}",
      journal = {\aap},
         year = 2014,
        month = apr,
       volume = {564},
          eid = {A92},
        pages = {A92},
          doi = {10.1051/0004-6361/201322377},
archivePrefix = {arXiv},
       eprint = {1309.0946},
 primaryClass = {astro-ph.EP},
       adsurl = {https://ui.adsabs.harvard.edu/abs/2014A&A...564A..92D}
}

@ARTICLE{2014Icar..239..118B,
       author = {{Berthier}, J. and {Vachier}, F. and {Marchis}, F. and {{\v{D}}urech}, J. and {Carry}, B.},
        title = "{Physical and dynamical properties of the main belt triple Asteroid (87) Sylvia}",
      journal = {\icarus},
         year = 2014,
        month = sep,
       volume = {239},
        pages = {118-130},
          doi = {10.1016/j.icarus.2014.05.046},
archivePrefix = {arXiv},
       eprint = {1407.1292},
 primaryClass = {astro-ph.EP},
       adsurl = {https://ui.adsabs.harvard.edu/abs/2014Icar..239..118B}
}

@ARTICLE{2014MNRAS.443.1802B,
       author = {{Bartczak}, P. and {Micha{\l}owski}, T. and {Santana-Ros}, T. and {Dudzi{\'n}ski}, G.},
        title = "{A new non-convex model of the binary asteroid 90 Antiope obtained with the SAGE modelling technique}",
      journal = {\mnras},
         year = 2014,
        month = sep,
       volume = {443},
       number = {2},
        pages = {1802-1809},
          doi = {10.1093/mnras/stu1247},
archivePrefix = {arXiv},
       eprint = {1406.6555},
 primaryClass = {astro-ph.EP},
       adsurl = {https://ui.adsabs.harvard.edu/abs/2014MNRAS.443.1802B}
}

@ARTICLE{2015A&A...578A..42R,
       author = {{Ryan}, E.~L. and {Mizuno}, D.~R. and {Shenoy}, S.~S. and {Woodward}, C.~E. and {Carey}, S.~J. and {Noriega-Crespo}, A. and {Kraemer}, K.~E. and {Price}, S.~D.},
        title = "{The kilometer-sized Main Belt asteroid population revealed by Spitzer}",
      journal = {\aap},
         year = 2015,
        month = jun,
       volume = {578},
          eid = {A42},
        pages = {A42},
          doi = {10.1051/0004-6361/201321375},
       adsurl = {https://ui.adsabs.harvard.edu/abs/2015A&A...578A..42R}
}

@ARTICLE{2015ApJ...814..117N,
       author = {{Nugent}, C.~R. and {Mainzer}, A. and {Masiero}, J. and {Bauer}, J. and {Cutri}, R.~M. and {Grav}, T. and {Kramer}, E. and {Sonnett}, S. and {Stevenson}, R. and {Wright}, E.~L.},
        title = "{NEOWISE Reactivation Mission Year One: Preliminary Asteroid Diameters and Albedos}",
      journal = {\apj},
         year = 2015,
        month = dec,
       volume = {814},
       number = {2},
          eid = {117},
        pages = {117},
          doi = {10.1088/0004-637X/814/2/117},
archivePrefix = {arXiv},
       eprint = {1509.02522},
 primaryClass = {astro-ph.EP},
       adsurl = {https://ui.adsabs.harvard.edu/abs/2015ApJ...814..117N}
}

@ARTICLE{2015Icar..256..101H,
       author = {{Hanu{\v{s}}}, J. and {Delbo'}, M. and {{\v{D}}urech}, J. and {Al{\'\i}-Lagoa}, V.},
        title = "{Thermophysical modeling of asteroids from WISE thermal infrared data - Significance of the shape model and the pole orientation uncertainties}",
      journal = {\icarus},
         year = 2015,
        month = aug,
       volume = {256},
        pages = {101-116},
          doi = {10.1016/j.icarus.2015.04.014},
archivePrefix = {arXiv},
       eprint = {1504.04199},
 primaryClass = {astro-ph.EP},
       adsurl = {https://ui.adsabs.harvard.edu/abs/2015Icar..256..101H}
}

@ARTICLE{2016A&A...585A...9L,
       author = {{Licandro}, J. and {Al{\'\i}-Lagoa}, V. and {Tancredi}, G. and {Fern{\'a}ndez}, Y.},
        title = "{Size and albedo distributions of asteroids in cometary orbits using WISE data}",
      journal = {\aap},
         year = 2016,
        month = jan,
       volume = {585},
          eid = {A9},
        pages = {A9},
          doi = {10.1051/0004-6361/201526866},
archivePrefix = {arXiv},
       eprint = {1510.02282},
 primaryClass = {astro-ph.EP},
       adsurl = {https://ui.adsabs.harvard.edu/abs/2016A&A...585A...9L}
}

@ARTICLE{2016A&A...591A..14A,
       author = {{Al{\'\i}-Lagoa}, V. and {Licandro}, J. and {Gil-Hutton}, R. and {Ca{\~n}ada-Assandri}, M. and {Delbo'}, M. and {de Le{\'o}n}, J. and {Campins}, H. and {Pinilla-Alonso}, N. and {Kelley}, M.~S.~P. and {Hanu{\v{s}}}, J.},
        title = "{Differences between the Pallas collisional family and similarly sized B-type asteroids}",
      journal = {\aap},
         year = 2016,
        month = jun,
       volume = {591},
          eid = {A14},
        pages = {A14},
          doi = {10.1051/0004-6361/201527660},
       adsurl = {https://ui.adsabs.harvard.edu/abs/2016A&A...591A..14A}
}

@ARTICLE{2016AJ....152...63N,
       author = {{Nugent}, C.~R. and {Mainzer}, A. and {Bauer}, J. and {Cutri}, R.~M. and {Kramer}, E.~A. and {Grav}, T. and {Masiero}, J. and {Sonnett}, S. and {Wright}, E.~L.},
        title = "{NEOWISE Reactivation Mission Year Two: Asteroid Diameters and Albedos}",
      journal = {\aj},
         year = 2016,
        month = sep,
       volume = {152},
       number = {3},
          eid = {63},
        pages = {63},
          doi = {10.3847/0004-6256/152/3/63},
archivePrefix = {arXiv},
       eprint = {1606.08923},
 primaryClass = {astro-ph.EP},
       adsurl = {https://ui.adsabs.harvard.edu/abs/2016AJ....152...63N}
}

@ARTICLE{2017A&A...599A..36H,
       author = {{Hanu{\v{s}}}, J. and {Marchis}, F. and {Viikinkoski}, M. and {Yang}, B. and {Kaasalainen}, M.},
        title = "{Shape model of asteroid (130) Elektra from optical photometry and disk-resolved images from VLT/SPHERE and Nirc2/Keck}",
      journal = {\aap},
         year = 2017,
        month = mar,
       volume = {599},
          eid = {A36},
        pages = {A36},
          doi = {10.1051/0004-6361/201629592},
archivePrefix = {arXiv},
       eprint = {1611.03632},
 primaryClass = {astro-ph.EP},
       adsurl = {https://ui.adsabs.harvard.edu/abs/2017A&A...599A..36H}
}

@ARTICLE{2017A&A...601A.114H,
       author = {{Hanu{\v{s}}}, J. and {Viikinkoski}, M. and {Marchis}, F. and {{\v{D}}urech}, J. and {Kaasalainen}, M. and {Delbo'}, M. and {Herald}, D. and {Frappa}, E. and {Hayamizu}, T. and {Kerr}, S. and {Preston}, S. and {Timerson}, B. and {Dunham}, D. and {Talbot}, J.},
        title = "{Volumes and bulk densities of forty asteroids from ADAM shape modeling}",
      journal = {\aap},
         year = 2017,
        month = may,
       volume = {601},
          eid = {A114},
        pages = {A114},
          doi = {10.1051/0004-6361/201629956},
archivePrefix = {arXiv},
       eprint = {1702.01996},
 primaryClass = {astro-ph.EP},
       adsurl = {https://ui.adsabs.harvard.edu/abs/2017A&A...601A.114H}
}

@ARTICLE{2017A&A...603A..55A,
       author = {{Al{\'\i}-Lagoa}, V. and {Delbo'}, M.},
        title = "{Sizes and albedos of Mars-crossing asteroids from WISE/NEOWISE data}",
      journal = {\aap},
         year = 2017,
        month = jul,
       volume = {603},
          eid = {A55},
        pages = {A55},
          doi = {10.1051/0004-6361/201629917},
archivePrefix = {arXiv},
       eprint = {1705.10263},
 primaryClass = {astro-ph.EP},
       adsurl = {https://ui.adsabs.harvard.edu/abs/2017A&A...603A..55A}
}

@ARTICLE{2017A&A...607A.117V,
       author = {{Viikinkoski}, M. and {Hanu{\v{s}}}, J. and {Kaasalainen}, M. and {Marchis}, F. and {{\v{D}}urech}, J.},
        title = "{Adaptive optics and lightcurve data of asteroids: twenty shape models and information content analysis}",
      journal = {\aap},
         year = 2017,
        month = nov,
       volume = {607},
          eid = {A117},
        pages = {A117},
          doi = {10.1051/0004-6361/201731456},
archivePrefix = {arXiv},
       eprint = {1708.05191},
 primaryClass = {astro-ph.EP},
       adsurl = {https://ui.adsabs.harvard.edu/abs/2017A&A...607A.117V}
}

@ARTICLE{2017AJ....154..168M,
       author = {{Masiero}, Joseph R. and {Nugent}, C. and {Mainzer}, A.~K. and {Wright}, E.~L. and {Bauer}, J.~M. and {Cutri}, R.~M. and {Grav}, T. and {Kramer}, E. and {Sonnett}, S.},
        title = "{NEOWISE Reactivation Mission Year Three: Asteroid Diameters and Albedos}",
      journal = {\aj},
         year = 2017,
        month = oct,
       volume = {154},
       number = {4},
          eid = {168},
        pages = {168},
          doi = {10.3847/1538-3881/aa89ec},
archivePrefix = {arXiv},
       eprint = {1708.09504},
 primaryClass = {astro-ph.EP},
       adsurl = {https://ui.adsabs.harvard.edu/abs/2017AJ....154..168M}
}

@ARTICLE{2018A&A...612A..85A,
       author = {{Al{\'\i}-Lagoa}, V. and {M{\"u}ller}, T.~G. and {Usui}, F. and {Hasegawa}, S.},
        title = "{The AKARI IRC asteroid flux catalogue: updated diameters and albedos}",
      journal = {\aap},
         year = 2018,
        month = may,
       volume = {612},
          eid = {A85},
        pages = {A85},
          doi = {10.1051/0004-6361/201731806},
archivePrefix = {arXiv},
       eprint = {1712.07496},
 primaryClass = {astro-ph.EP},
       adsurl = {https://ui.adsabs.harvard.edu/abs/2018A&A...612A..85A}
}

@ARTICLE{2018Icar..309..297H,
       author = {{Hanu{\v{s}}}, J. and {Delbo'}, M. and {{\v{D}}urech}, J. and {Al{\'\i}-Lagoa}, V.},
        title = "{Thermophysical modeling of main-belt asteroids from WISE thermal data}",
      journal = {\icarus},
         year = 2018,
        month = jul,
       volume = {309},
        pages = {297-337},
          doi = {10.1016/j.icarus.2018.03.016},
archivePrefix = {arXiv},
       eprint = {1803.06116},
 primaryClass = {astro-ph.EP},
       adsurl = {https://ui.adsabs.harvard.edu/abs/2018Icar..309..297H}
}

@ARTICLE{2019Sci...364..268W,
       author = {{Watanabe}, S. and {Hirabayashi}, M. and {Hirata}, N. and {Hirata}, Na. and {Noguchi}, R. and {Shimaki}, Y. and {Ikeda}, H. and {Tatsumi}, E. and {Yoshikawa}, M. and {Kikuchi}, S. and {Yabuta}, H. and {Nakamura}, T. and {Tachibana}, S. and {Ishihara}, Y. and {Morota}, T. and {Kitazato}, K. and {Sakatani}, N. and {Matsumoto}, K. and {Wada}, K. and {Senshu}, H. and {Honda}, C. and {Michikami}, T. and {Takeuchi}, H. and {Kouyama}, T. and {Honda}, R. and {Kameda}, S. and {Fuse}, T. and {Miyamoto}, H. and {Komatsu}, G. and {Sugita}, S. and {Okada}, T. and {Namiki}, N. and {Arakawa}, M. and {Ishiguro}, M. and {Abe}, M. and {Gaskell}, R. and {Palmer}, E. and {Barnouin}, O.~S. and {Michel}, P. and {French}, A.~S. and {McMahon}, J.~W. and {Scheeres}, D.~J. and {Abell}, P.~A. and {Yamamoto}, Y. and {Tanaka}, S. and {Shirai}, K. and {Matsuoka}, M. and {Yamada}, M. and {Yokota}, Y. and {Suzuki}, H. and {Yoshioka}, K. and {Cho}, Y. and {Tanaka}, S. and {Nishikawa}, N. and {Sugiyama}, T. and {Kikuchi}, H. and {Hemmi}, R. and {Yamaguchi}, T. and {Ogawa}, N. and {Ono}, G. and {Mimasu}, Y. and {Yoshikawa}, K. and {Takahashi}, T. and {Takei}, Y. and {Fujii}, A. and {Hirose}, C. and {Iwata}, T. and {Hayakawa}, M. and {Hosoda}, S. and {Mori}, O. and {Sawada}, H. and {Shimada}, T. and {Soldini}, S. and {Yano}, H. and {Tsukizaki}, R. and {Ozaki}, M. and {Iijima}, Y. and {Ogawa}, K. and {Fujimoto}, M. and {Ho}, T. -M. and {Moussi}, A. and {Jaumann}, R. and {Bibring}, J. -P. and {Krause}, C. and {Terui}, F. and {Saiki}, T. and {Nakazawa}, S. and {Tsuda}, Y.},
        title = "{Hayabusa2 arrives at the carbonaceous asteroid 162173 Ryugu{\textemdash}A spinning top-shaped rubble pile}",
      journal = {Science},
         year = 2019,
        month = apr,
       volume = {364},
       number = {6437},
        pages = {268-272},
          doi = {10.1126/science.aav8032},
       adsurl = {https://ui.adsabs.harvard.edu/abs/2019Sci...364..268W}
}

@ARTICLE{2019pdss.data....3H,
       author = {{Herald}, D. and {Frappa}, E. and {Gault}, D. and {Hayamizu}, T. and {Kerr}, S. and {Moore}, J. and {Giacchini}, B.},
        title = "{Small Bodies Occultations Bundle V3.0}",
      journal = {NASA Planetary Data System},
         year = 2019,
        month = dec,
        pages = {3},
          doi = {10.26033/ap0g-wf63},
       adsurl = {https://ui.adsabs.harvard.edu/abs/2019pdss.data....3H}
}

@ARTICLE{2020A&A...633A..65H,
       author = {{Hanu{\v{s}}}, J. and {Vernazza}, P. and {Viikinkoski}, M. and {Ferrais}, M. and {Rambaux}, N. and {Podlewska-Gaca}, E. and {Drouard}, A. and {Jorda}, L. and {Jehin}, E. and {Carry}, B. and {Marsset}, M. and {Marchis}, F. and {Warner}, B. and {Behrend}, R. and {Asenjo}, V. and {Berger}, N. and {Bronikowska}, M. and {Brothers}, T. and {Charbonnel}, S. and {Colazo}, C. and {Coliac}, J. -F. and {Duffard}, R. and {Jones}, A. and {Leroy}, A. and {Marciniak}, A. and {Melia}, R. and {Molina}, D. and {Nadolny}, J. and {Person}, M. and {Pejcha}, O. and {Riemis}, H. and {Shappee}, B. and {Sobkowiak}, K. and {Sold{\'a}n}, F. and {Suys}, D. and {Szakats}, R. and {Vantomme}, J. and {Birlan}, M. and {Berthier}, J. and {Bartczak}, P. and {Dumas}, C. and {Dudzi{\'n}ski}, G. and {{\v{D}}urech}, J. and {Castillo-Rogez}, J. and {Cipriani}, F. and {Fetick}, R. and {Fusco}, T. and {Grice}, J. and {Kaasalainen}, M. and {Kryszczynska}, A. and {Lamy}, P. and {Michalowski}, T. and {Michel}, P. and {Santana-Ros}, T. and {Tanga}, P. and {Vachier}, F. and {Vigan}, A. and {Witasse}, O. and {Yang}, B.},
        title = "{(704) Interamnia: a transitional object between a dwarf planet and a typical irregular-shaped minor body}",
      journal = {\aap},
         year = 2020,
        month = jan,
       volume = {633},
          eid = {A65},
        pages = {A65},
          doi = {10.1051/0004-6361/201936639},
archivePrefix = {arXiv},
       eprint = {1911.13049},
 primaryClass = {astro-ph.EP},
       adsurl = {https://ui.adsabs.harvard.edu/abs/2020A&A...633A..65H}
}

%
%

\onecolumn 
\begin{appendix}

\section{Extra figures of the method}\label{sec:app_met_figs}

\begin{figure}[H]
\centering
\import{figures}{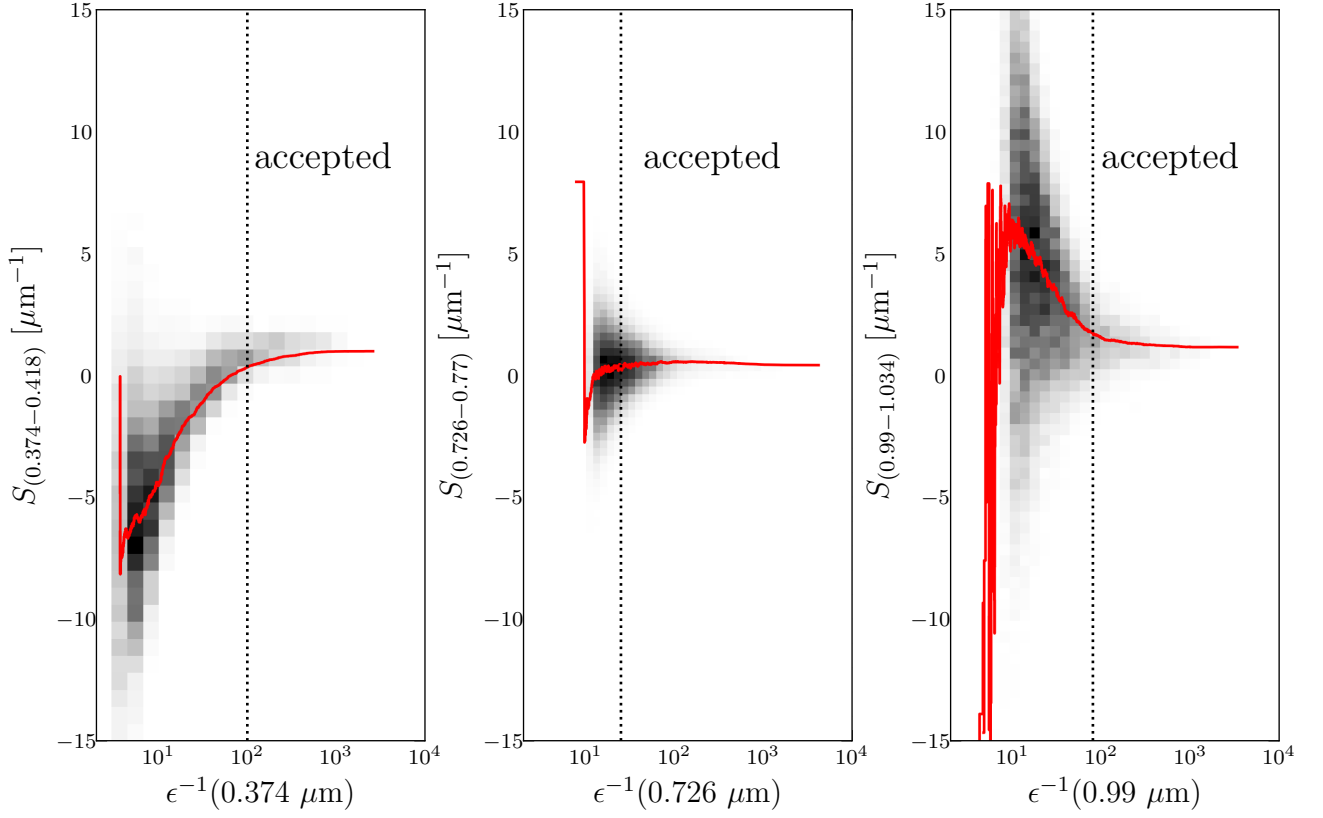}
\caption{Distribution of the spectral slope between two consecutive bands versus the uncertainty. We selected as a example the two first (left panel), two central (middle panel), and the two last bands (right panel). The red solid line is a running median of the distribution. The black dashed vertical lines indicate the thresholds we used to clean the sample from low SNR spectra that could be affecting the clustering process.}
\label{fig:dat:SNR_cut}
\end{figure}

\begin{figure}
\centering
\import{figures}{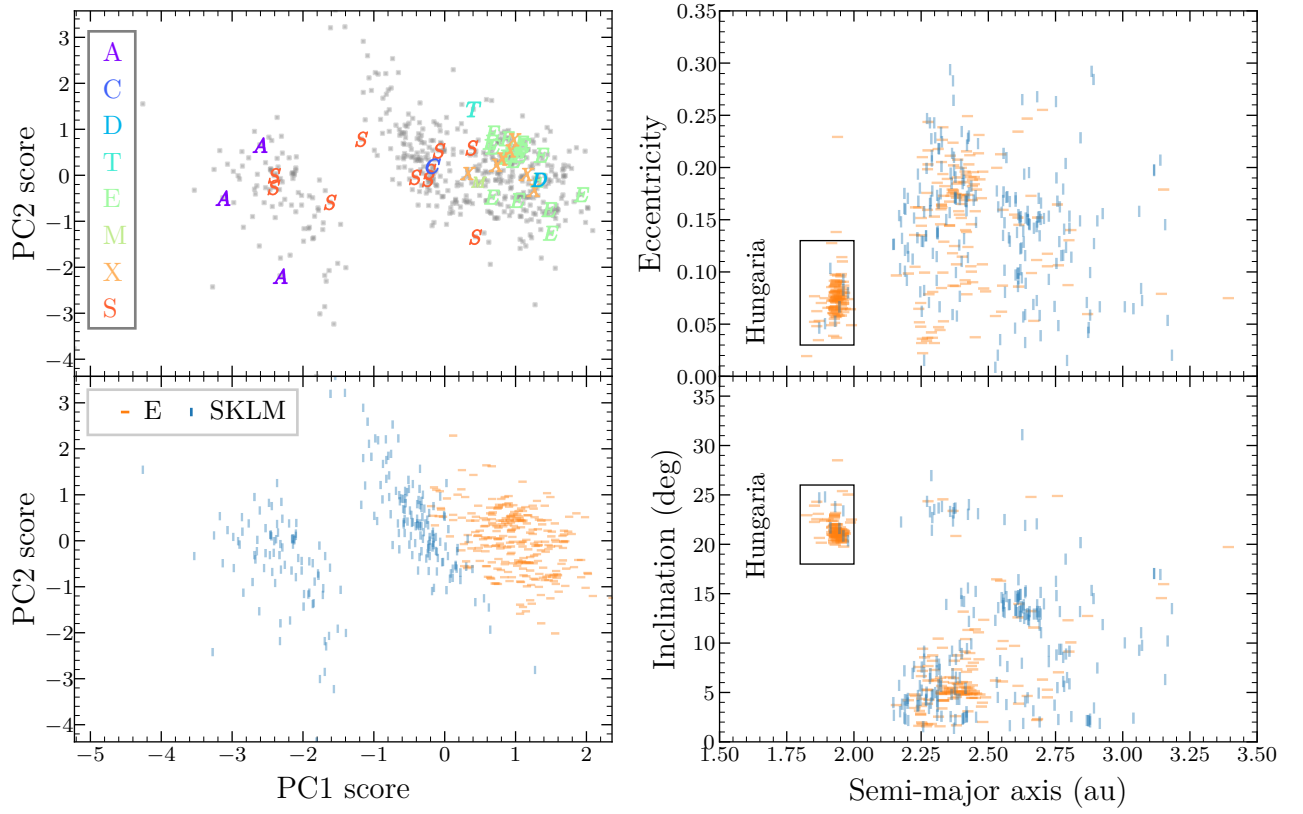}
\caption{Same as Fig. \ref{fig:met:0ft} but over the E-complex and using three features: log$_{10}$($p_v$), $\Delta S_{0.5-0.7-0.9}$, and $S_{0.55-0.9}$. Now, the blue cluster is the outlier asteroids polluting this complex, and the yellow cluster is named E taxon. In the bottom panels, we highlighted the Hungarian population.}
\label{fig:met:E}
\end{figure}

\begin{figure}
\centering
\import{figures}{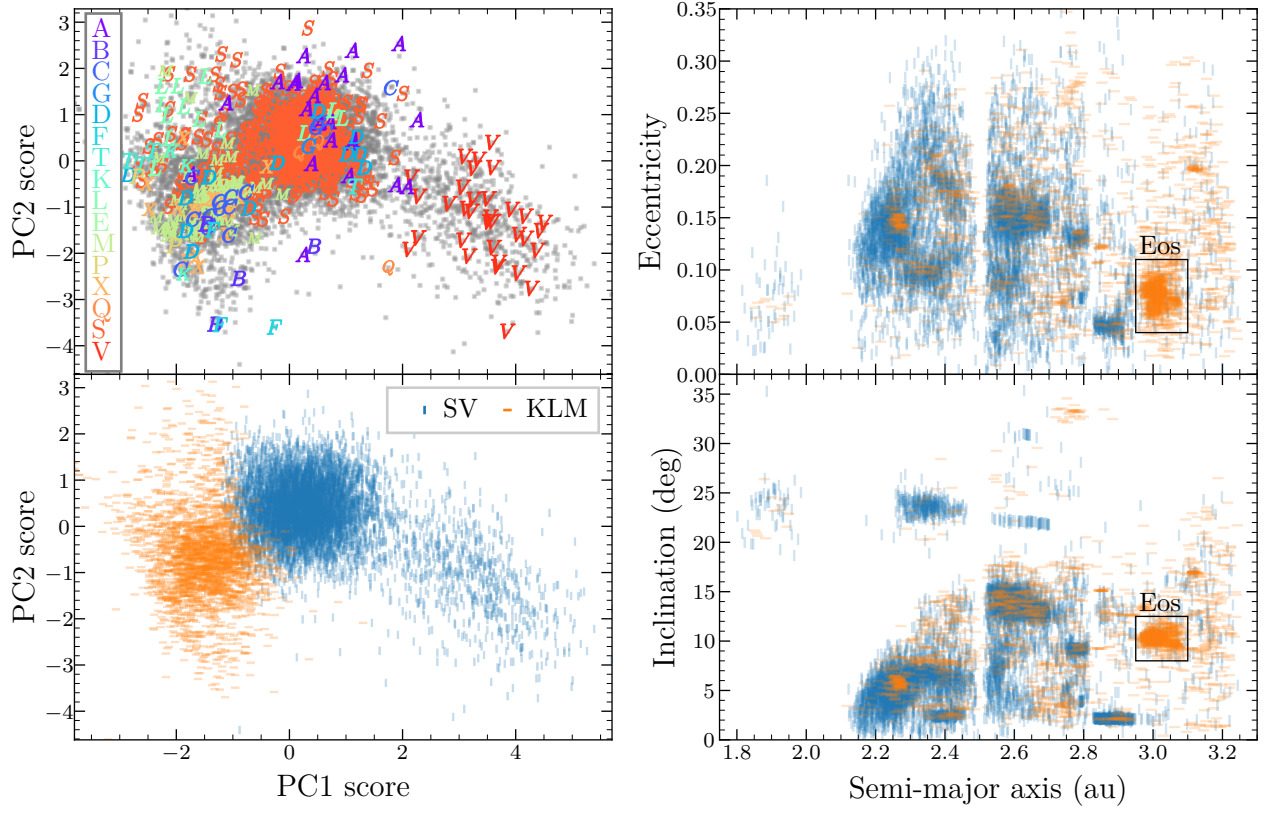}
\caption{Same as Fig. \ref{fig:met:0ft} but over the SKLM-complex formed by S, KLM complexes and outliers from E-complex. In the PCA we used three features: log$_{10}$($p_v$), $\Delta S_{0.5-0.7-0.9}$, and $S_{0.55-0.9}$. Now, the blue cluster is the SV complex, and the orange cluster is the KLM complex. In the bottom panels, we highlighted the Eos collisional family.}
\label{fig:met:SKLM}
\end{figure}

\begin{figure}
\centering
\import{figures}{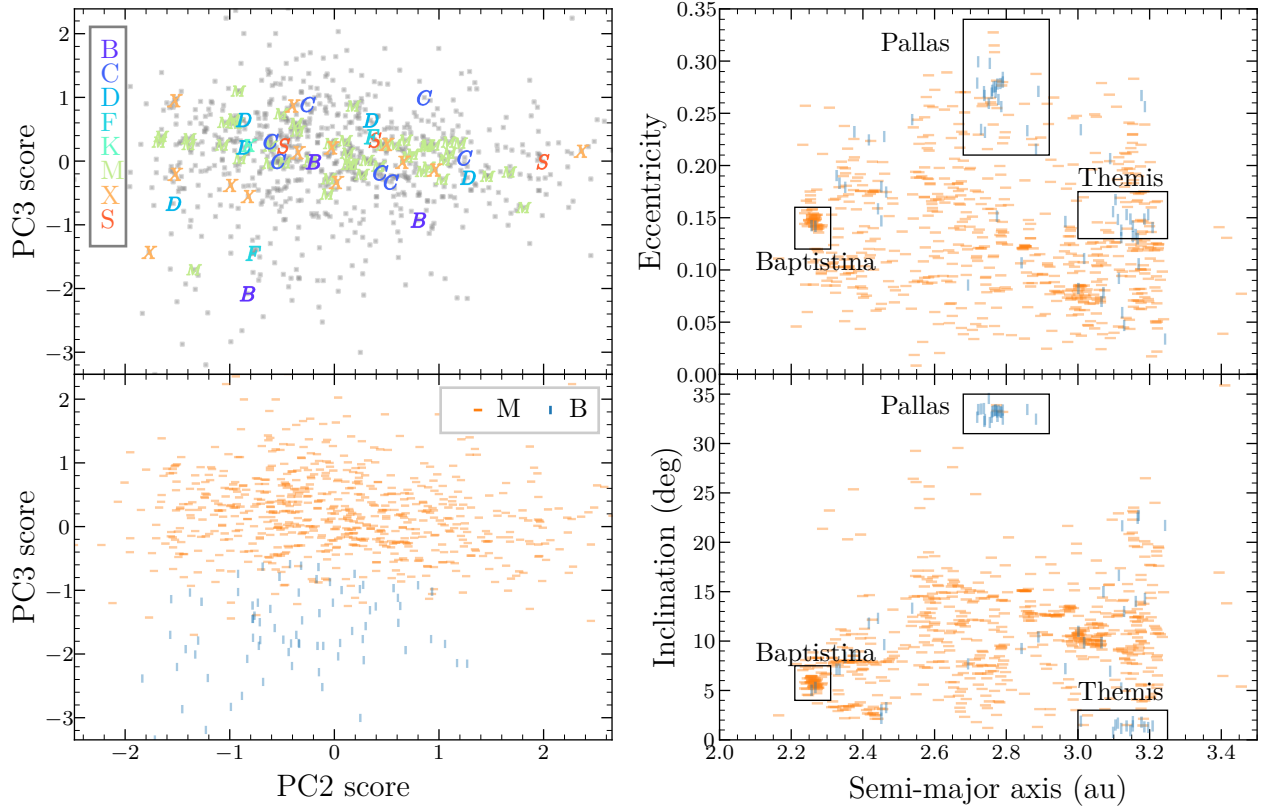}
\caption{Same as Fig. \ref{fig:met:0ft} but over the C-complex and using  three features: log$_{10}$($p_v$), $S_{0.5-0.9}$, and $\Delta S_{0.46-0.55-0.73}$. We used this time PC3 vs PC2 as it is where the blue cluster is more differentiable. Now, the orange cluster is named M taxon, and the blue cluster is joined with the B taxon defined in Fig. \ref{fig:res:BF} (see text). In the bottom panels, we highlighted the collisional families known to be related to M-type asteroids (Baptistina) and B-types (Pallas and Themis).}
\label{fig:met:BM}
\end{figure}

\begin{figure}
\centering
\import{figures}{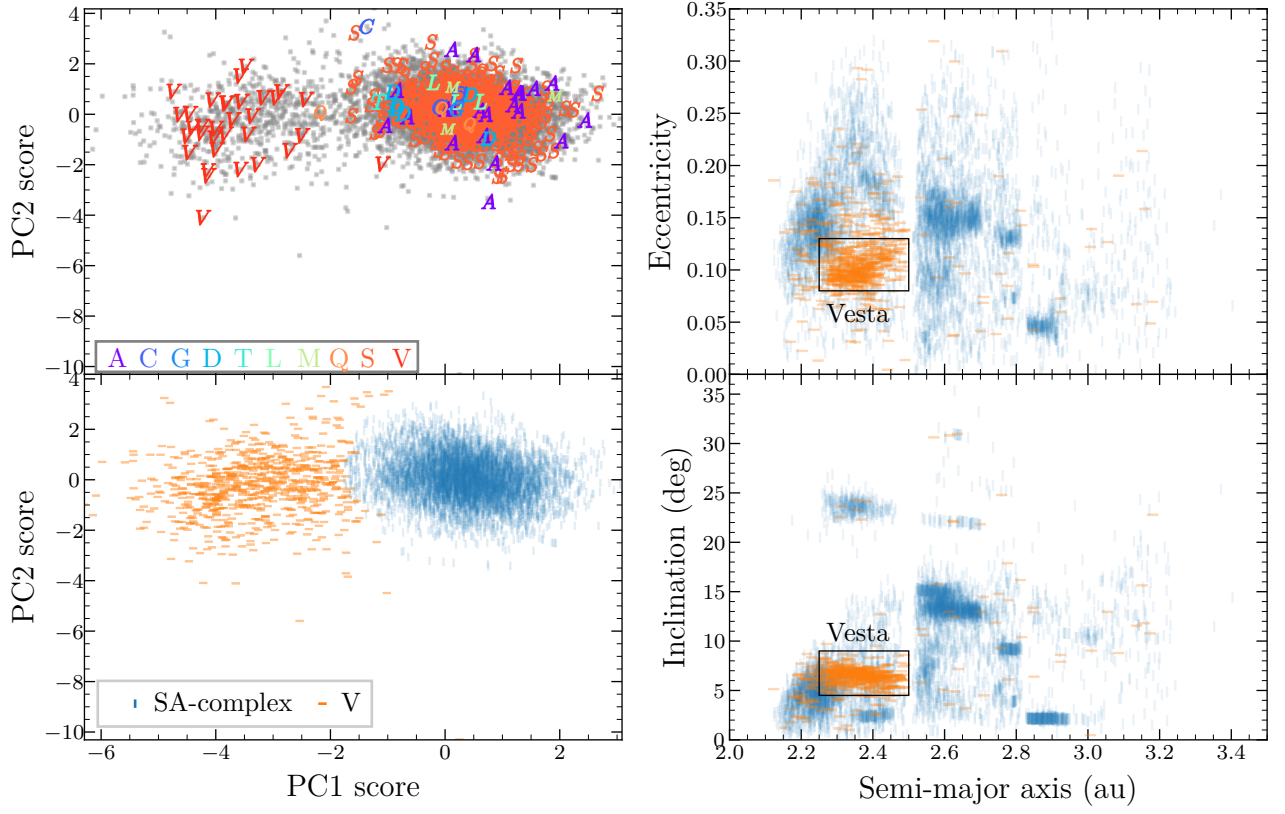}
\caption{Same as Fig. \ref{fig:met:0ft} but over the SV-complex and using  three features: log$_{10}$($p_v$), $S_{0.81-0.9}$, and med(${0.81-0.9}$). The blue cluster is named SA-complex, and the orange cluster is named the V taxon. In the right panels, we highlighted the collisional family known to be related to V-type asteroids, Vesta.}
\label{fig:met:SV}
\end{figure}

\FloatBarrier

\section{Extra tables for results}

\begin{table}[H]
\caption{
Number of objects classified per taxon and dynamical population}\label{tab:app_res:perc}
\centering
\begin{tabular}{ c r  r  r  r  r  r  r  r}  
\hline\hline
  \multirow{3}{*}{Taxon}   & \multicolumn{7}{c}{Dynamical Population (total)} \\
  \cmidrule{2-8}
      & Hungaria &  IMB   &  MMB   & OMB    & Cybele & Hilda & JT \\
      & (174)    & (4317) & (4542) & (4351) & (126)  & (105) & (180) \\
\hline
A & 6   & 69   & 29   & 24   & 0  & 0   &  0\\
    B & 0   & 16   & 53   & 199  & 2  & 0   &  1\\
    C & 1   & 398  & 885  & 1544 & 27 & 13  &  23\\
    G & 0   & 37   & 116  & 137  & 5  & 0   &  0\\
    D & 0   & 44   & 73   & 309  & 49 & 71  &  148\\
    F & 2   & 49   & 38   & 36   & 1  & 0   &  0\\
    K & 6   & 180  & 236  & 919  & 3  & 5   &  1\\
    L & 4   & 48   & 103  & 62   & 1  & 0   &  0\\
    E & 105 & 120  & 25   & 3    & 1  & 0   &  0\\
    M & 2   & 174  & 203  & 293  & 5  & 2   &  1\\
    P & 0   & 24   & 120  & 231  & 29 & 14  &  6\\
    S & 48  & 2489 & 2632 & 580  & 3  & 0   &  0\\
    V & 0   & 669  & 29   & 14   & 0  & 0   &  0\\
\hline
\end{tabular}
\end{table}

\begin{table}[H]
\caption{
Descriptors of the albedo distribution per taxon}\label{tab:app_res:perc}
\centering
\begin{tabular}{ c | r  r  r  r | r | r  r  r  r}  
\hline
 Tax & min & $P_{001}$ & $P_{003}$ & $P_{016}$ & $P_{050}$ & $P_{084}$ & $P_{097}$ & $P_{099}$ & max \\
 \hline
 P  & 0.02  & 0.02  & 0.03  & 0.03  & 0.04  & 0.06  & 0.08  & 0.09  & 0.09  \\
 F  & 0.03  & 0.03  & 0.03  & 0.04  & 0.05  & 0.06  & 0.06  & 0.07  & 0.07  \\
 G  & 0.03  & 0.03  & 0.03  & 0.04  & 0.05  & 0.06  & 0.08  & 0.08  & 0.09  \\
 C  & 0.01  & 0.03  & 0.03  & 0.04  & 0.05  & 0.07  & 0.08  & 0.09  & 0.1  \\
 D  & 0.02  & 0.03  & 0.03  & 0.04  & 0.06  & 0.07  & 0.08  & 0.09  & 0.09  \\
 B  & 0.04  & 0.04  & 0.04  & 0.05  & 0.06  & 0.1  & 0.13  & 0.15  & 0.16  \\
 M  & 0.07  & 0.09  & 0.09  & 0.1  & 0.13  & 0.17  & 0.21  & 0.23  & 0.24  \\
 K  & 0.02  & 0.08  & 0.09  & 0.1  & 0.13  & 0.15  & 0.19  & 0.21  & 0.29  \\
 L  & 0.12  & 0.13  & 0.13  & 0.14  & 0.17  & 0.21  & 0.25  & 0.27  & 0.28 \\
 S  & 0.1   & 0.13  & 0.15  & 0.18  & 0.22  & 0.27  & 0.33  & 0.36  & 0.44 \\
 A  & 0.1   & 0.11  & 0.13  & 0.19  & 0.23  & 0.27  & 0.35  & 0.37  & 0.38 \\
 V  & 0.02  & 0.14  & 0.17  & 0.22  & 0.28  & 0.35  & 0.45  & 0.52  & 0.67 \\
 E  & 0.24  & 0.25  & 0.27  & 0.36  & 0.52  & 0.64  & 0.8  & 0.87  & 1.43 \\ 
 \hline
\end{tabular}
\end{table}

\FloatBarrier
\section{References of visual albedos and diameters}\label{sec:app_ref}

\begin{table*}
    \caption{Data references used for albedos and diameters.}
    \centering
    \begin{tabular}{p{18cm}}
    \\
    \hline
\text{\cite{2013A&A...554A..71A,2016A&A...591A..14A,2018A&A...612A..85A,2020A&A...638A..84A}}; \text{\cite{2017A&A...603A..55A}}; \text{\cite{2014MNRAS.443.1802B}}; \text{\cite{2013ApJ...773...22B}}; \text{\cite{2014Icar..239..118B,2023A&A...671A.151B}}; \text{\cite{2021MNRAS.502.4981C}}; \text{\cite{1999Icar..140...53C}}; \text{\cite{2004PhDT.......371D}}; \text{\cite{2009P&SS...57..259D}}; \text{\cite{2014A&A...564A..92D}}; \text{\cite{2011Icar..214..652D}}; \text{\cite{2013A&A...555A..15F}}; \text{\cite{2011ApJ...742...40G, 2012ApJ...744..197G, 2012ApJ...759...49G}}; \text{\cite{2013PASJ...65...34H}}; \text{\cite{2013Icar..226.1045H,2015Icar..256..101H,2017A&A...599A..36H,2017A&A...601A.114H,2018Icar..309..297H,2020A&A...633A..65H}}; \text{\cite{2019pdss.data....3H}}; \text{\cite{2012MNRAS.423.2587H}}; \text{\cite{2022PSJ.....3...56H}}; \text{\cite{2021AJ....162...40J}}; \text{\cite{2016ApJ...817L..22L}}; \text{\cite{2016A&A...585A...9L}}; \text{\cite{2007Icar..186..152M,2007Icar..186..126M}}; \text{\cite{2011ApJ...743..156M, 2012ApJ...760L..12M}}; \text{\cite{2006Icar..185...39M,2012Icar..221.1130M}}; \text{\cite{2023A&A...670A..52M}}; \text{\cite{2011ApJ...741...68M,2012ApJ...759L...8M,2014ApJ...791..121M,2017AJ....154..168M,2020PSJ.....1....5M,2021PSJ.....2..162M}}; \text{\cite{2021A&A...652A.141M}}; \text{\cite{2022PSJ.....3...30M}}; \text{\cite{2015ApJ...814..117N,2016AJ....152...63N}}; \text{\cite{2012Icar..221..365P}}; \text{\cite{2016Sci...353.1008R}}; \text{\cite{2015A&A...578A..42R}}; \text{\cite{2010AJ....140..933R,2011AJ....141..186R}}; \text{\cite{2008Icar..193...20S}}; \text{\cite{2019Sci...364..252S}}; \text{\cite{2002AJ....123.1056T}}; \text{\cite{1999Icar..140...17T}};  \text{\cite{2010AJ....140..770T}}; \text{\cite{2011PASJ...63.1117U}}; \text{\cite{2020NatAs...4..136V,2021A&A...654A..56V}}; \text{\cite{2017A&A...607A.117V}};  \text{\cite{2019Sci...364..268W}};\text{\cite{2020A&A...641A..80Y}}\\
    \hline
    \end{tabular}
    \label{tab:alb_ref}
\end{table*}

\end{appendix}
\end{document}